\documentclass{article}

\usepackage{arxiv}

\usepackage[utf8]{inputenc} 
\usepackage[T1]{fontenc}    
\usepackage{hyperref}       
\usepackage{url}            
\usepackage{booktabs}       
\usepackage{amsfonts}       
\usepackage{nicefrac}       
\usepackage{microtype}      
\usepackage{lipsum}		
\usepackage{graphicx}
\usepackage{natbib}
\usepackage{doi}
\usepackage{makecell}
\usepackage{amssymb}
\usepackage{amsmath}
\usepackage[ruled,vlined]{algorithm2e}

\title{Ensemble ToT of LLMs and Its Application to Automatic Grading System \\for Supporting Self-Learning}

\author{ \href{https://orcid.org/0009-0009-2073-8831}{\includegraphics[scale=0.06]{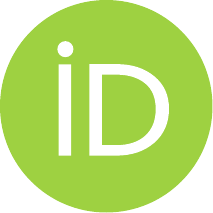}\hspace{1mm}Yuki Ito}\\
Graduate School of Informatics \\
Kyoto University \\
Yoshidahonmachi, Sakyo-ku\\
Kyoto-shi, Kyoto 606-8501, Japan\\ 
\texttt{ito.yuki.w26@kyoto-u.jp}
	\And
	\href{https://orcid.org/0000-0003-3430-9244}{\includegraphics[scale=0.06]{orcid.pdf}\hspace{1mm}Qiang Ma} \\
    Graduate School of Science and Technology\\ 
    Kyoto Institute of Technology\\
    Matsugasakihashikamicho, Sakyo-ku,\\
    Kyoto 606-8585, Japan\\
\texttt{qiang@kit.ac.jp}\\
}

\renewcommand{\shorttitle}{Ensemble ToT for Automatic Grading}

\hypersetup{
pdftitle={Ensemble ToT of LLMs and Its Application to Automatic Grading System for Supporting Self-Learning},
pdfsubject={cs.CL, cs.AI},
pdfauthor={Yuki Ito, Qiang Ma},
pdfkeywords={Large Language Models, EdTech, Ensemble, ToT.},
}

\begin{document}
\maketitle

\begin{abstract}
    Providing students with detailed and timely grading feedback is essential for self-learning.
    While existing LLM-based grading systems are promising, most of them rely on one single model, which limits their performance.
    To address this, we propose \textbf{Ensemble Tree-of-Thought (ToT)}, a framework that enhances LLM outputs by integrating multiple models. Using this framework, we develop a grading system.
    Ensemble ToT follows three steps: (1) analyzing LLM performance, (2) generating candidate answers, and (3) refining them into a final result.
    Based on this, our grading system first evaluates the grading tendencies of LLMs, then generates multiple results, and finally integrates them via a simulated debate.
    Experimental results demonstrate our approach's ability to provide accurate and explainable grading by effectively coordinating multiple LLMs.
\end{abstract}


\section{Introduction}\label{sec-introduction}
Since the release of ChatGPT in 2022, generative AI has gained significant popularity, with many students increasingly adopting it as a learning aid for various purposes.
A survey on students' use of AI for learning identified several common applications, such as brainstorming, generating ideas, summarizing readings, and analyzing large datasets \citep{voicesonai}.
Among these, AI's ability to provide personalized and immediate learning support is particularly valued.
Notably, some students in the survey reported using AI to obtain grading and its explanations on their completed homework.
They found these grading comments helpful for enhancing their depth of thinking and understanding.
This highlights the importance of providing students with detailed, point-by-point feedback on their homework to facilitate self-learning.

However, generative AI responses are not always reliable, raising concerns about the potential for misleading feedback.
While feedback from human teachers is ideal for guiding student self-learning, it is often impractical for teachers to provide individualized and immediate support to every student.
In this context, automatic grading systems emerge as a practical solution.
These systems evaluate student answers and deliver timely grading comments, bridging the gap without placing additional burdens on teachers.

Before this work, several studies have proposed automatic grading methods.
For example, Gobbo et al. introduced GradeAid, a framework that evaluates student answers by extracting both lexical and semantic features to determine grading scores \citep{gradeaid}.
Similarly, Huang et al. developed a system to grade reading comprehension answers \citep{direct}.
Their system not only evaluates student answers but also provides auto-generated hints to guide students toward correct answers.

Despite these advances, existing automated grading systems often fall short of effectively supporting self-learning.
Many systems lack the ability to provide high-quality, detailed feedback necessary for guiding students' learning processes.
For instance, GradeAid does not explain \textit{why} an answer is incorrect or incomplete.
This leaves students without clear guidance on how to improve.
While some methods attempt to generate more detailed explanations about their grading results, the quality remains inadequate.
For example, Huang et al. say about their system that their auto-generated feedback quality is ``still far from satisfactory."
These deficiencies hinder the effectiveness of the systems for self-learning.

To overcome these challenges, integrating multiple LLMs into grading systems is a promising approach.
Existing methods typically rely on a single LLM, which may struggle to provide consistently accurate and reliable grading comments.
In contrast, a previous study demonstrated that using ensembles of pre-trained BERT models for scoring tasks significantly improves prediction accuracy \citep{ensembleoflm}.
This highlights the advantages of combining multiple models.
Inspired by this insight, we advocate for leveraging multi-LLM systems to address the limitations of single-model approaches.
However, an effective way to combine multiple LLMs, especially language generation models, has not yet been well studied.

\begin{figure}[!tb]
    \centering
    \includegraphics[width=\textwidth]{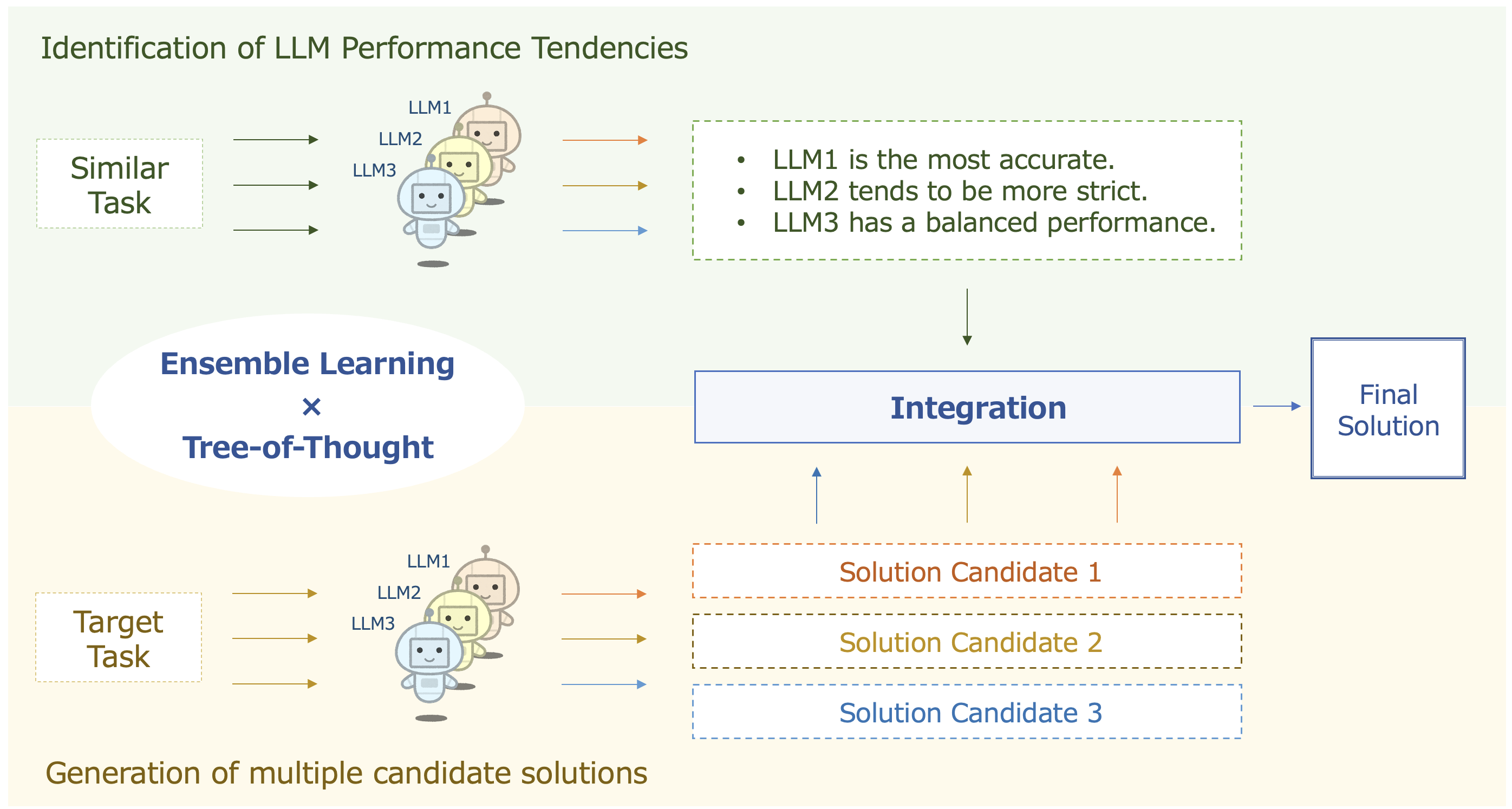}
    \caption{\textbf{Overview of Ensemble ToT Framework}: The framework integrates ensemble learning techniques with the Tree-of-Thought (ToT) approach. It identifies individual LLM performance tendencies and synthesizes multiple candidate solutions generated by LLMs into a single refined result.}
    \label{fig-ensembletot}
\end{figure}

In this paper, we propose a novel framework to utilize multiple LLMs called \textbf{Ensemble ToT}  and its application, a grading system named \textbf{Graders by Ensemble ToT (GET)}.
Ensemble ToT combines the strengths of multiple language generation models by integrating ensemble learning techniques with the Tree-of-Thought (ToT) approach.
Fig.\,\ref{fig-ensembletot} shows the overview of this framework.
It operates in three key steps:
\begin{enumerate}
    \item \textbf{Identification of LLM performance tendencies}: Inspired by ensemble learning principles, we evaluate the performance of individual LLMs on tasks similar to the target problem. This helps us identify the strengths and tendencies of each LLM for the given task.
    \item \textbf{Generation of multiple candidate solutions}: Drawing on the ToT framework, multiple LLMs independently generate candidate solutions for the task.
    \item \textbf{Integration of the solutions}: These candidate solutions are synthesized into a single, refined result, leveraging insights from the performance tendencies.
\end{enumerate}

\begin{figure}[!tb]
    \centering
    \includegraphics[width=\textwidth]{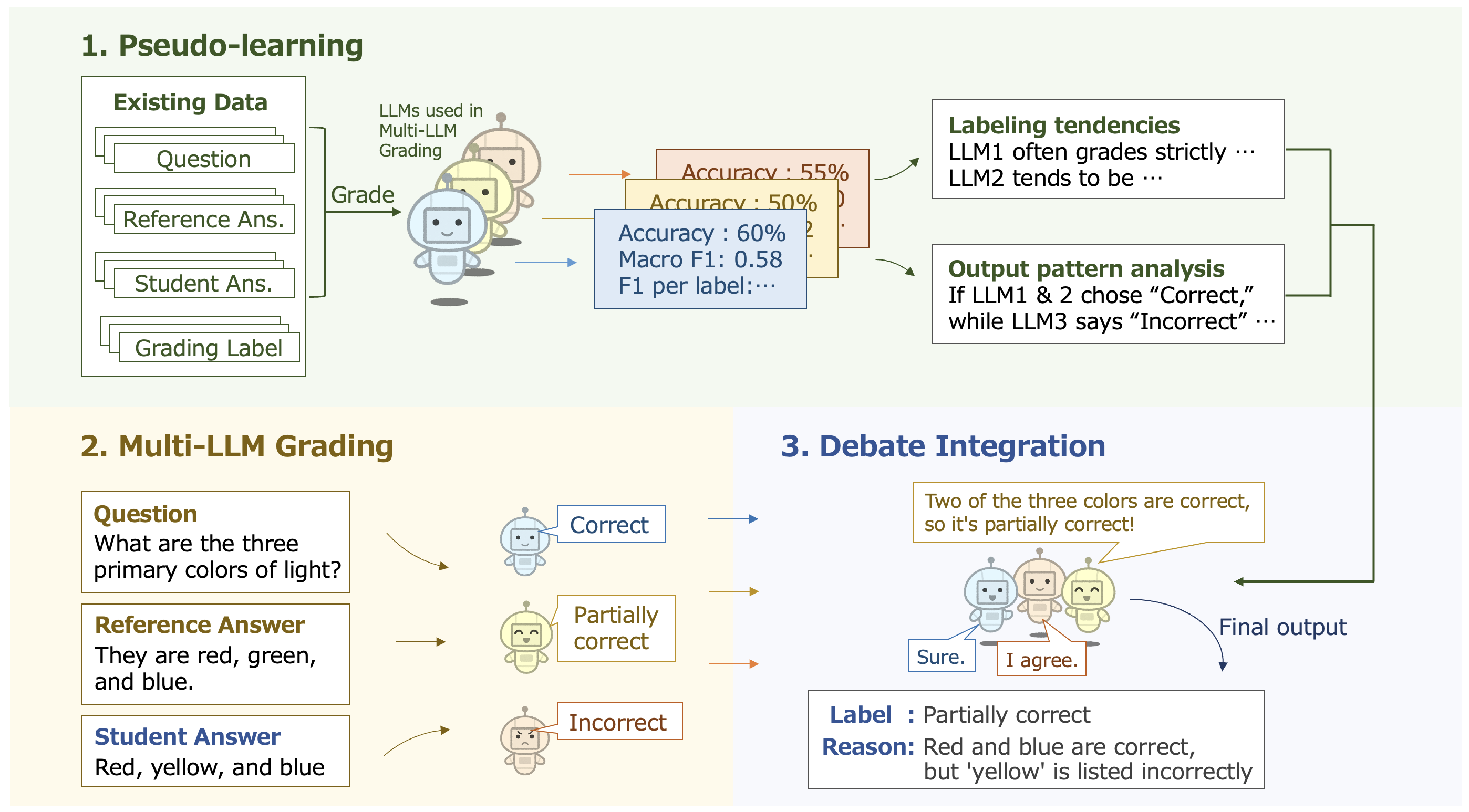}
    \caption{\textbf{Process Diagram of GET}: The system consists of three stages: \textit{pseudo-learning}, \textit{multi-LLM grading}, and \textit{debate Integration}. This enables accurate grading by taking advantage of the characteristics of multiple models.}
    \label{fig-overview}
\end{figure}

The GET system leverages the Ensemble ToT framework to deliver highly accurate grading and its detailed explanations — both of which are difficult to achieve using one single LLM.
Fig.\,\ref{fig-overview} provides the overview of the GET system process flow.
Following the Ensemble ToT structure, it operates in three main steps: \textit{pseudo-learning}, \textit{multi-LLM grading}, \textit{debate integration}.

In \textit{pseudo-learning}, three LLMs separately grade past student answers and determine their grading labels.
By analyzing the results of this task, grading capabilities and tendencies of LLMs are identified.
Next, in the \textit{multi-LLM grading} step, the three LLMs separately generate grading candidates of student answers.
They get questions, reference answers, and student answers as the inputs and output grading labels along with their reasons.
Finally, in the \textit{debate integration} step, an LLM integrates \textit{multi-LLM grading} results into the single, refined grading conclusion.
This integration is accomplished by simulating a debate script between the grader LLMs.
\textit{Pseudo-learning}'s outcome is utilized in this integration phase to guide debate outputs.

By providing students with grading reasons accurately and speedily, the GET system enhances the explainability of grading outcomes, helping students understand and correct their mistakes.
Furthermore, by combining the output results from multiple LLMs, the system compensates for the weaknesses of individual models.
This advancement enables more accurate and balanced grading, accompanied by clear and valid reasoning for student answers.
It is expected to enhance student self-learning while minimizing the workload for educators.

The key contributions of this work are as follows:
\begin{itemize}
    \item Propose Ensemble ToT, a novel framework to collaborate multiple LLMs effectively. It contains three main steps.
          \begin{itemize}
              \item Analysis of LLMs performance tendencies inspired by ensemble learning.
              \item Generation of several candidate solutions with multiple LLMs based on the Tree-of-Thought framework.
              \item Integration of candidate solutions considering the performance tendencies.
          \end{itemize}
    \item Propose an automated grading system, GET, which is constructed based on the Ensemble ToT framework.
    \item Evaluate the grading performance of the GET system on existing datasets.  By comparing it to several baselines, we guarantee its effectiveness.
\end{itemize}

\section{Related Work}\label{sec-relatedwork}
\subsection{Methods to Collaborate Multiple LLMs}

Recent advancements in large language models (LLMs) have explored strategies to leverage the collective strengths of multiple models, enhancing performance and reliability across various tasks.

For example, Zhang et al. introduce the Chain-of-Agent (CoA) framework, which utilizes multi-agent collaboration in natural language \citep{CoA}.
In CoA, multiple LLM agents communicate to process different segments of text.
Then, a manager agent synthesizes their contributions into the final output.
With each agent processing short contexts, CoA can handle longer contextual input more effectively than conventional methods.
It has demonstrated significant improvements (up to 10\%) in tasks such as question answering, summarization, and code completion, when compared to strong baselines like retrieval augmented generation (RAG), full-context models, and multi-agent LLMs.

The LLM-Synergy framework \citep{yang2023one} proposes techniques like boosting-based weighted majority voting and cluster-based dynamic model selection to enhance accuracy in medical QA datasets.
This approach achieves state-of-the-art results, including 38.01\% accuracy on MedMCQA, 96.36\% on PubMedQA, and 38.13\% on MedQA-USMLE, outperforming other methods across all three datasets.

Additionally, the Formal Debate framework (FORD) \citep{ford} introduces a structured three-stage debate process to facilitate effective collaboration among LLMs.
While the integrated results are not always optimal, the study demonstrates the potential of debates to improve performance even when there are inconsistencies between LLMs' outputs.

These methods highlight the potential of multi-LLM collaboration in addressing challenges related to performance and reliability.

Based on the insights provided by these studies, our study explores a new approach to achieving multi-LM cooperation.
Unlike the aforementioned frameworks, our system introduces two key elements: \textit{pseudo-learning}, which pre-evaluates the performance trends of LLMs, and \textit{debate integration}, which synthesizes LLMs' outputs through discussions.
Through these mechanisms, our system aims to effectively address performance and reliability challenges.

\subsection{Automated Short Answer Grading}
The application of a LLM in education has also received significant attention.
Automated short answer grading (ASAG) field is no exception.
ASAG targets the grading of student answers to open-ended questions, which are typically a few sentences long.
In this section, we introduce such technologies, by categorizing existing grading methods into two primary approaches: \textbf{score or label prediction} and \textbf{detailed feedback generation}.

\subsubsection{Grading Methods for Scores and Labels Prediction}
Numerous studies have explored grading methods that focus on assigning scores or labels to student answers.
For example, Gobbo et al. propose GradeAid, a framework to grade short student answers.
It gives numeric scores to student answers by extracting lexical and semantic features \citep{gradeaid} from them.
Lexical features are calculated using TF-IDF, while semantic features are derived using a BERT-Cross Encoder.
These features are concatenated into a single vector and processed through a regression model to predict the final grading score.
GradeAid demonstrates effective grading across various datasets, while its performance consistency depends on the specific questions and datasets used.
Another approach utilizes ensembles of pre-trained BERT models for scoring tasks \citep{ensembleoflm}.
This study shows that combining multiple models can improve score prediction accuracy.

While BERT models are used in various grading tasks, GPT models are also becoming popular these days.
For example, Chang et al. employ GPT-3.5 and GPT-4 for scoring and label classification (pass/fail) of Finnish student answers \citep{finnishgrading}.
This research found that GPT-4 performs well in the grading tasks, especially in the one-shot setting.
But they also admit the necessity of further performance improvement before introducing this kind of technology to the real education situation.

Automated grading can reduce not only the burden on teachers but also the subjectivity inherent in manual grading.
Gobrecht et al. indicate that automated scoring methods can enhance fairness and equity in evaluations \citep{beyondhumansubjectivity}.
They fine-tuned a transformer-based scoring system and evaluated student answers which already assigned grades by human graders.
Then, human graders grade the same answers again.
Finally, these results are compared with the past grading to evaluate the grading consistency.
This experiment reveals that the model has less deviation from the past grading results than the human re-graders.

\subsubsection{Grading Methods for Detailed Feedback Generation}
Some methods generate detailed grading reasons alongside scores and labels to enhance transparency and usability.
For example, Huang et al. proposes a method that automatically evaluates answers using a fine-tuned GPT model and generates feedback to guide students to the correct answers \citep{direct}.
We have also proposed a technique to automatically generate grading labels and reasons by combining a single LLM and a grammatical structure analysis technique \citep{icetc}.

In addition, Lee et al.'s study \citep{cotscoring} uses a prompt construction method called Chain-of-Thought (CoT) to perform reasoned automatic grading with GPT-3.5 and GPT-4 \citep{cotscoring}.
CoT prompting is a way to improve LLM outputs by encouraging step-by-step reasoning.
In this system, the LLM generates evidence that supports its grading decisions before assigning a label (e.g., Proficient, Developing, or Beginning).
This approach enables GPT-4 to provide explanations for its assigned labels, improving interpretability.
Another study investigated GPT-4's potential for grading middle school level answers \citep{sasgpt4}.
Their method generates both grading scores and rationale paragraphs, by referring to the question, student answer, scoring rubrics, and additional information where applicable.
GPT-4 achieved a quadratic weighted kappa of 0.677 across 10 questions.
While this experimental result shows a high potential of auto-grading by the model, it is also mentioned that grading outcomes varied across subjects (e.g., biology, English), highlighting the instability.

Some studies propose grading frameworks that can work with various types of LLMs.
For example, Jordan et al. developed an automated grading framework: FreeText \citep{freetext}.
It gets a question, grading criteria, and a student answer, then generates personalized feedback for the student.
In this framework, users can freely select a grading model.
Fateen et al. propose another system, ASAS-F-RAG, which integrates retrieval-augmented generation (RAG) with automated grading \citep{beyondscores}.
When grading student answers, this method retrieves the top 3 to 5 most similar examples from past grading data using the ColBERT retriever.
These examples are given to the LLMs as grading examples.
This method achieves accurate generation of scores, labels (e.g., Correct, Partially Correct, or Incorrect), and feedback.
They also employ several LLMs in their experiments, indicating that their system achieves accurate grading without depending on a specific LLM.

\subsubsection{Dataset Contributions for Automated Grading}
While many studies focus on label and score prediction, fewer studies address the generation of grading reasons.
This discrepancy stems from the limited availability of datasets that include detailed grading reasons.
To address this, several works have introduced new datasets to support research on ASAG:

\begin{itemize}
    \item SAF Dataset \citep{safdataset}: Filighera et al. construct the SAF dataset collected from college-level network communication classes. It includes questions, reference answers, student answers, grading labels, scores, and grader feedback. They demonstrated the dataset's utility by fine-tuned models like T5-base and mT5-base to jointly predict scores, labels, and feedback.

    \item EngSAF Dataset \citep{iunderstandwhy}: Aggarwal et al. construct another dataset: EngSAF. It has a similar structure to the SAF, but it expands the scope with a larger data size and broader topics across various courses. Experiments using EngSAF involved both fine-tuned models (e.g., LLaMA-2-13B-chat, Mistral-7B-Instruct) and zero-shot evaluations with GPT-3.5-turbo. These models generated output labels and feedback by comparing student answers to reference answers.
\end{itemize}

These datasets are crucial for advancing the development of systems that prioritize reason generation alongside grading.

\subsubsection{Comparison to Our Approach}

Tables \ref{table-outputs} and \ref{table-techniques} summarize the features and techniques of the reviewed methods, contrasting them with our system.

\begin{table}[!htb]
    \centering
    \caption{Outputs of Each System}
    \label{table-outputs}
    \renewcommand{\arraystretch}{1.3}
    \begin{tabular}{lccc}
        \hline
        \textbf{Method/System}                              & \makecell{\textbf{Label Pred.}} & \makecell{\textbf{Score Pred.}} & \textbf{Reason Gen.} \\ \hline
        GradeAid \citep{gradeaid}                           & ×                               & \checkmark                      & ×                    \\
        Finnish Grading \citep{finnishgrading}              & \checkmark                      & \checkmark                      & ×                    \\
        BERT Ensemble \citep{ensembleoflm}                  & ×                               & \checkmark                      & ×                    \\
        Reduce Subjectivity \citep{beyondhumansubjectivity} & ×                               & \checkmark                      & ×                    \\
        CoT Scoring \citep{cotscoring}                      & \checkmark                      & ×                               & \checkmark           \\
        FreeText\citep{freetext}                            & ×                               & ×                               & \checkmark           \\
        ASAS-F-RAG \citep{beyondscores}                     & \checkmark                      & \checkmark                      & \checkmark           \\
        GPT-4 Grading \citep{sasgpt4}                       & ×                               & \checkmark                      & \checkmark           \\
        SAF Baselines \citep{safdataset}                    & \checkmark                      & \checkmark                      & \checkmark           \\
        EngSAF Baselines \citep{iunderstandwhy}             & \checkmark                      & \checkmark                      & \checkmark           \\
        DIRECT \citep{direct}                               & \checkmark                      & ×                               & \checkmark           \\
        Our Previous System \citep{icetc}                   & \checkmark                      & ×                               & \checkmark           \\

        \textbf{GET} (Our System)                           & \checkmark                      & ×                               & \checkmark           \\ \hline
    \end{tabular}
\end{table}

\begin{table}[!htb]
    \centering
    \caption{Techniques Used in Each Method}
    \label{table-techniques}
    \renewcommand{\arraystretch}{1.3}
    \begin{tabular}{lcccc}
        \hline
        \textbf{Method/System}                              & \makecell{\textbf{Fine-Tuning}} & \makecell{\textbf{Few-Shot}} & \makecell{\textbf{CoT/}\textbf{ToT}} & \makecell{\textbf{Multi-LLM} \\\textbf{Collab.}} \\ \hline
        GradeAid \citep{gradeaid}                           & \checkmark                      & ×                            & ×                                    & ×                            \\
        Finnish Grading \citep{finnishgrading}              & ×                               & \checkmark                   & ×                                    & ×                            \\
        BERT Ensemble \citep{ensembleoflm}                  & \checkmark                      & ×                            & ×                                    & \checkmark                   \\
        Reduce Subjectivity \citep{beyondhumansubjectivity} & \checkmark                      & ×                            & ×                                    & ×                            \\
        CoT Scoring \citep{cotscoring}                      & ×                               & \checkmark                   & \checkmark                           & ×                            \\
        FreeText \citep{cotscoring}                         & ×                               & ×                            & ×                                    & ×                            \\
        ASAS-F-RAG \citep{beyondscores}                     & ×                               & \makecell{\checkmark }       & ×                                    & ×                            \\
        GPT-4 Grading \citep{sasgpt4}                       & ×                               & \checkmark                   & \checkmark                           & ×                            \\
        SAF Baselines \citep{safdataset}                    & \checkmark                      & \checkmark                   & ×                                    & ×                            \\
        EngSAF  Baselines \citep{iunderstandwhy}            & \checkmark                      & ×                            & ×                                    & ×                            \\
        DIRECT \citep{direct}                               & \checkmark                      & ×                            & ×                                    & ×                            \\
        Our Previous System \citep{icetc}                   & \checkmark                      & ×                            & ×                                    & ×                            \\

        \textbf{GET} (Our System)                           & ×                               & \checkmark                   & \checkmark                           & \checkmark                   \\ \hline
    \end{tabular}
\end{table}

As shown in Table \ref{table-outputs}, our system focuses on label prediction and reason generation, excluding score prediction.
This design choice aligns with our ultimate goal: providing effective feedback to support student self-learning rather than determining official grades.

Table \ref{table-techniques} highlights how our system distinguishes itself from prior methods.
Unlike approaches that rely on a single model, our method leverages multiple LLMs to enhance grading accuracy.
This is achieved through a combination of ensemble learning and the Tree of Thought (ToT) approach.
By collaborating multiple LLMs, we can compensate for the weaknesses of individual models.
This results in more consistent and balanced grading outcomes, without relying on a specific LLM.

Note that the ToT approach extends the Chain of Thought (CoT) methodology.
While CoT simulates a linear flow of thought in LLMs, ToT explores multiple alternatives simultaneously.
It branches the reasoning into a tree-like structure.
This branching enables more nuanced grading decisions and improves accuracy.

Furthermore, our system avoids the computationally expensive process of fine-tuning.
Instead of training a new model to combine model outputs (as in traditional ensemble learning), we focus on identifying the performance tendencies of various LLMs.
These identified tendencies are coupled with few-shot learning and the ToT framework, allowing us to achieve high performance without the overhead of fine-tuning.
By omitting fine-tuning, we reduce the time and computational resources required.
This enables faster and more efficient system deployment.

\section{Proposed Framework and System}\label{sec-proposedmethod}

This section first introduces the core idea, Ensemble ToT.
Then, it defines the problem that the GET system is designed to solve.
Finally, it provides a detailed explanation of the system's overall structure.

\subsection{Ensemble ToT}\label{sec-ensembletot}
Inspired by ensemble learning and Tree-of-Thought (ToT) \citep{ToT}, we propose a novel framework, Ensemble ToT, for utilizing multiple LLMs.

Ensemble learning is a machine learning technique that combines the predictions from multiple models to improve accuracy \citep{ensemblelearning}.
One common method, Stacking \citep{stacking}, involves training multiple models on the same dataset and then training a new model to combine their outputs.
This new model generally produces higher-quality results by leveraging the strengths of the individual models.

In contrast, ToT is a framework designed to help LLMs solve problems by breaking them down into a series of intermediate thought steps.
For each step, ToT guides the LLM to generate multiple candidate solutions. These candidates are then evaluated using heuristics or voting mechanisms to select the most promising options.
By iteratively exploring multiple reasoning paths, ToT organizes the problem-solving process into a tree-like structure, allowing LLMs to explore various paths simultaneously.

Ensemble ToT builds upon Stacking and ToT.
First, it identifies the performance tendencies of LLMs on tasks similar to the target problem.
This phase mirrors the training phase in Stacking. However, we do not actually train the LLMs, since they have large size, and fine-tuning them is resource-intensive.
To address this, we design an alternative process called \textit{pseudo-learning}, as described in Section \ref{subsec-pseudo-learning}.

Second, based on the ToT framework, Ensemble ToT generates multiple candidate solutions using several LLMs.
While the original ToT uses a single LLM to generate multiple candidates, Ensemble ToT utilizes different LLMs for each candidate.
This increases the diversity of the candidates, leading to a wider range of solutions.
In the GET system, this is implemented in the \textit{multi-LLM grading} phase, detailed in Section \ref{subsec-multi-llm-grading}.

Finally, Ensemble ToT combines the candidate solutions while considering the identified performance tendencies of the LLMs.
This process corresponds both to ensemble learning, which combines the outputs of multiple models through a newly trained model, and to ToT, which merges thought candidates using heuristics or voting mechanisms.
In Ensemble ToT, an LLM combines the performance tendencies and candidate solutions to select the best possible solution.
The GET system implements this process in the \textit{debate integration} phase, as explained in Section \ref{subsec-debateintegration}.

The following section provides a detailed explanation of the GET system, which implements the Ensemble ToT framework.

\subsection{Problem Definition}
The GET system's objective is to grade short student answers.
To define this problem, the symbols used are explained in Table \ref{tab-terminology}.

\begin{table}[!b]
    \centering
    \caption{Definition of Symbols}
    \label{tab-terminology}
    \renewcommand{\arraystretch}{1.3}
    \begin{tabular}{ll}
        \hline
        \textbf{Symbol}                                                       & \textbf{Definition}                    \\ \hline $Q = \{q_1, q_2, \ldots, q_m\}$ & Set of questions for students \\
        $R = \{r_1, r_2, \ldots, r_m\}$                                       & Set of reference answers for Q         \\
        $S_i = \{s_{i_1}, s_{i_2}, \ldots, s_{i_n}\}$                         & Set of $n$ students' answers for $q_i$ \\
        $GL = \{\text{Correct}, \text{Partially correct}, \text{Incorrect}\}$ & Set of grading labels                  \\
        $GR$                                                                  & Set of grading reasons                 \\
        $gl_{i_j} \in GL$                                                     & Grading label for  $s_{i_j}$           \\
        $gr_{i_j}$                                                            & Grading reason for  $s_{i_j}$          \\ \hline
    \end{tabular}
\end{table}

The problem addressed by the GET system is to grade student answers based on the question content and the corresponding reference answers.
Specifically, for the input tuple $(q_i, r_i, s_{i_j})$, the system predicts a pair consisting of a grading label and a grading reason, $(gl_{i_j}, gr_{i_j})$.

Mathematically, this task can be represented as follows:
\begin{equation}
    (gl_{i_j}, gr_{i_j}) = GET(q_i, r_i, s_{i_j})
\end{equation}

The GET system tackles this task through three steps: \textit{pseudo-learning}, \textit{multi-LLM grading}, and \textit{debate integration}.

\subsection{Pseudo-Learning}\label{subsec-pseudo-learning}

\textit{Pseudo-learning} aims to evaluate the capabilities and grading tendencies of LLMs employed in the system.
It is conducted before grading student answers.
The process comprises two steps: \textit{grading past-data} and \textit{tendencies identification}.

\subsubsection{Grading Past-Data}
In the \textit{grading past-data} step, each LLM independently grades answers from an existing dataset $D'$, such as historical student answers with grading results.
Then it computes the LLMs' performance metrics.

Formally, each LLM predicts grading label and reason pairs $(gl'_{i_j}, gr'_{i_j})$ based on the input tuples $(q'_i, r'_i, s'_{i_j}) \in D'$.
The predicted grading labels (denoted as $pl'_{i_j}$) are then evaluated against ground truth $gl'_{i_j}$ using performance metrics, including accuracy and macro F1-score.

Accuracy is defined as the proportion of accurately classified samples to the total number of samples:

\begin{equation}
    \text{Accuracy} = \frac{\text{Number of Accurate Predictions}}{\text{Total Number of Predictions}}
    \label{eq-acc}
\end{equation}

The macro F1-score is calculated as the arithmetic mean of F1-scores for all three grading labels in $GL$, ensuring equal weight for each class, regardless of its frequency.
For each label $gl$, the F1-score is defined as:
\begin{equation}
    F1_{gl} = \frac{2 \cdot \text{Precision}_{gl} \cdot \text{Recall}_{gl}}{\text{Precision}_{gl} + \text{Recall}_{gl}}
    \label{eq-singlef1}
\end{equation}
where:
\begin{equation}
    \text{Precision}_{gl} = \frac{\text{TP}_{gl}}{\text{TP}_{gl} + \text{FP}_{gl}}, \quad
    \text{Recall}_{gl} = \frac{\text{TP}_{gl}}{\text{TP}_{gl} + \text{FN}_{gl}}
    \label{eq-precisionrecall}
\end{equation}
\begin{itemize}
    \item \textbf{True Positives ($\text{TP}_{gl}$)}: The number of samples predicted as belonging to grading label $gl$ and having $gl$ as their true label.
    \item \textbf{False Positives ($\text{FP}_{gl}$)}: The number of samples predicted as belonging to grading label $gl$ but having true labels from other classes.
    \item \textbf{False Negatives ($\text{FN}_{gl}$)}: The number of samples with a true label of $gl$ but predicted as belonging to another grading label.
\end{itemize}

The macro F1-score is then computed as:
\begin{equation}
    \text{Macro F1} = \frac{1}{|GL|} \sum_{gl \in GL} F1_{gl}.
    \label{eq-macrof1}
\end{equation}
Throughout this study, these metrics are calculated using Scikit-learn \citep{scikit-learn}.

In the \textit{grading past-data} step, we focus solely on the evaluation of grading label prediction.
Grading reason evaluation is omitted because assessing natural language is complex and time-consuming.
This omission simplifies the process and allows for quicker implementation when additional LLMs will be used in the system.

\subsubsection{Tendencies Identification}
\textit{Tendencies identification} analyzes the strengths, weaknesses, and grading biases of each LLM based on the performance metrics calculated in the previous step.
The analysis is conducted from two perspectives:

\begin{itemize}
    \item \textbf{Labeling tendencies compared to other LLMs} (e.g., an LLM excels at identifying errors but is overly strict).
    \item \textbf{Output pattern analysis for the most likely label prediction} (e.g., if LLM1 and LLM2 assign \textit{Correct} while only LLM3 assigns \textit{Partially correct}, the ground truth is \textit{Partially correct} in many cases).
\end{itemize}

\begin{figure}[!htb]
    \centering
    \includegraphics[width=0.75\textwidth]{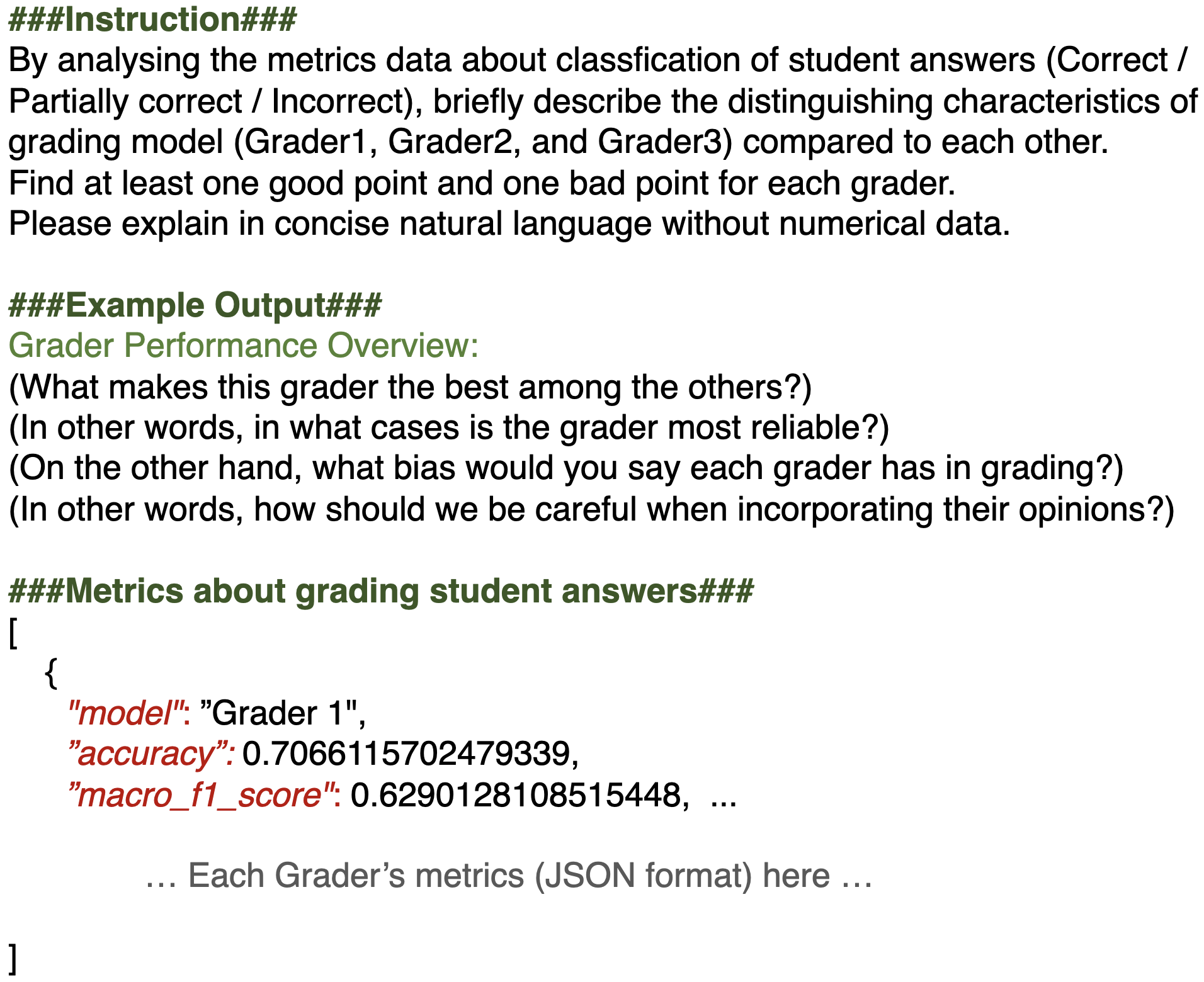}
    \caption{\textbf{The prompt used for analyzing labeling tendencies}: This figure illustrates the prompt provided to the LLM during the labeling tendencies analysis. It includes three-class classification performance metrics in JSON format, along with detailed instructions and the expected output.}
    \label{fig-extractgraderfeature}
\end{figure}

For labeling tendencies, an LLM interprets the numeric metrics and explains its findings in natural language.
The prompt is shown in Fig.\,\ref{fig-extractgraderfeature}.
Through this process, numerical data are transformed into text form.
LLMs are generally more effective at processing natural language than numerical data.
This transformation is expected to ensure the results of \textit{pseudo-learning} can be utilized more effectively in the subsequent processes.

For output pattern analysis, the system uses Algorithm \ref{alg-mostlikelylabel}.
It determines the most likely grading label for a given combination of LLM output labels by leveraging the \textit{grading past-data} results and the data distributions.
The algorithm works as follows:
\begin{enumerate}
    \item \textbf{Count occurrences}: For each student answer, count the occurrences of LLM output label combination $(pl'_1, pl'_2, pl'_3)$ and their associated ground truth labels $gl'$,  where $pl'_i \in GL$ is the predicted label of LLM $i$, and $gl' \in GL$ is the ground truth label.
    \item \textbf{Calculate ratios}: For each label combination, calculate the ratio of observed label counts to the expected label counts from the prior probabilities.
    \item \textbf{Assign most likely labels}: Output all labels whose ratios exceed a predefined threshold as most likely labels. If no label exceeds the threshold, the label with the highest frequency of occurrence is output.
\end{enumerate}

{\linespread{1.3}\selectfont
\begin{algorithm}[!htb]
    \SetAlgoLined
    \KwIn{Data $\{(pl'_1, pl'_2, pl'_3, gl')\}$, Prior Probabilities $\{P(\text{Label})\}$}
    \KwOut{Likely Label Results $\{(Combination, MostLikelyLabel) | \; Combination \in GL^3\}$}
    \vspace{2mm}

    \textbf{Initialize} $CombinationCounter \gets \{\}$, \\ $GroundTruthMap \gets \{\}$, \\ $LikelyLabelResults \gets \{\}$\;
    \vspace{2mm}
    \tcc{1. Count Occurrences and Collect Ground Truth Labels for Each Combination}
    \ForEach{$(pl'_1, pl'_2, pl'_3, gl') \in \text{Data}$}{
        $Combination \gets (pl'_1, pl'_2, pl'_3)$\;
        $CombinationCounter[Combination]++$\;
        $GroundTruthMap[Combination][gl']++$\;
    }
    \vspace{2mm}
    \tcc{2. Calculate Ratios of Ground Truth to Prior Probabilities}
    \ForEach{$(Combination, C\_Count) \in CombinationCounter.item()$}{
    $Ratios \gets \{\}$\;
    \vspace{2mm}
    \ForEach{$\text{Label, L\_Count} \in \text{GroundTruthMap[Combination].item()}$}{
        $Ratios[\text{Label}] \gets \frac{\text{L\_Count}}{P(\text{Label}) \cdot \text{C\_Count}}$\;
    }
    \vspace{2mm}
    \tcc{3. Select Labels Exceeding the Threshold}
    $SelectedLabels \gets \{\text{Labels where } Ratios[\text{Label}] > \text{Threshold}\}$\;
    \vspace{2mm}
    \uIf{$SelectedLabels \neq \emptyset$}{
        $LikelyLabelResults[Combination] \gets SelectedLabels$\;
    }
    \Else{
    \tcc{If all labels are less than the threshold, select the most frequent label}
    $MostFrequentLabel \gets \text{argmax}_{\text{Label}} \; GroundTruthMap[Combination][\text{Label}]$\;
    $LikelyLabelResults[Combination] \gets \{MostFrequentLabel\}$\;
    }
    }
    \Return{Likely Label Results}\;
    \caption{Determine Most Likely Label}
    \label{alg-mostlikelylabel}
\end{algorithm}}

The prior probability used in the algorithm is the percentage of the number of labels in the dataset.

\subsection{Multi-LLM Grading}\label{subsec-multi-llm-grading}

The \textit{multi-LLM grading} is the phase to generate multiple grading result candidates of the actual student answers.

This task can be formally defined as follows:
Given a question, reference answer, and student answer ($q_i$, $r_i$, $s_{i_j}$) as the input, each LLM predicts the grading label and reason ($gl_{i_j}$, $gr_{i_j}$).
This process finally generates three patterns of the predicted label and reason.

\begin{figure}[!htb]
    \centering
    \begin{minipage}[b]{0.55\textwidth}
        \centering
        \includegraphics[width=\textwidth]{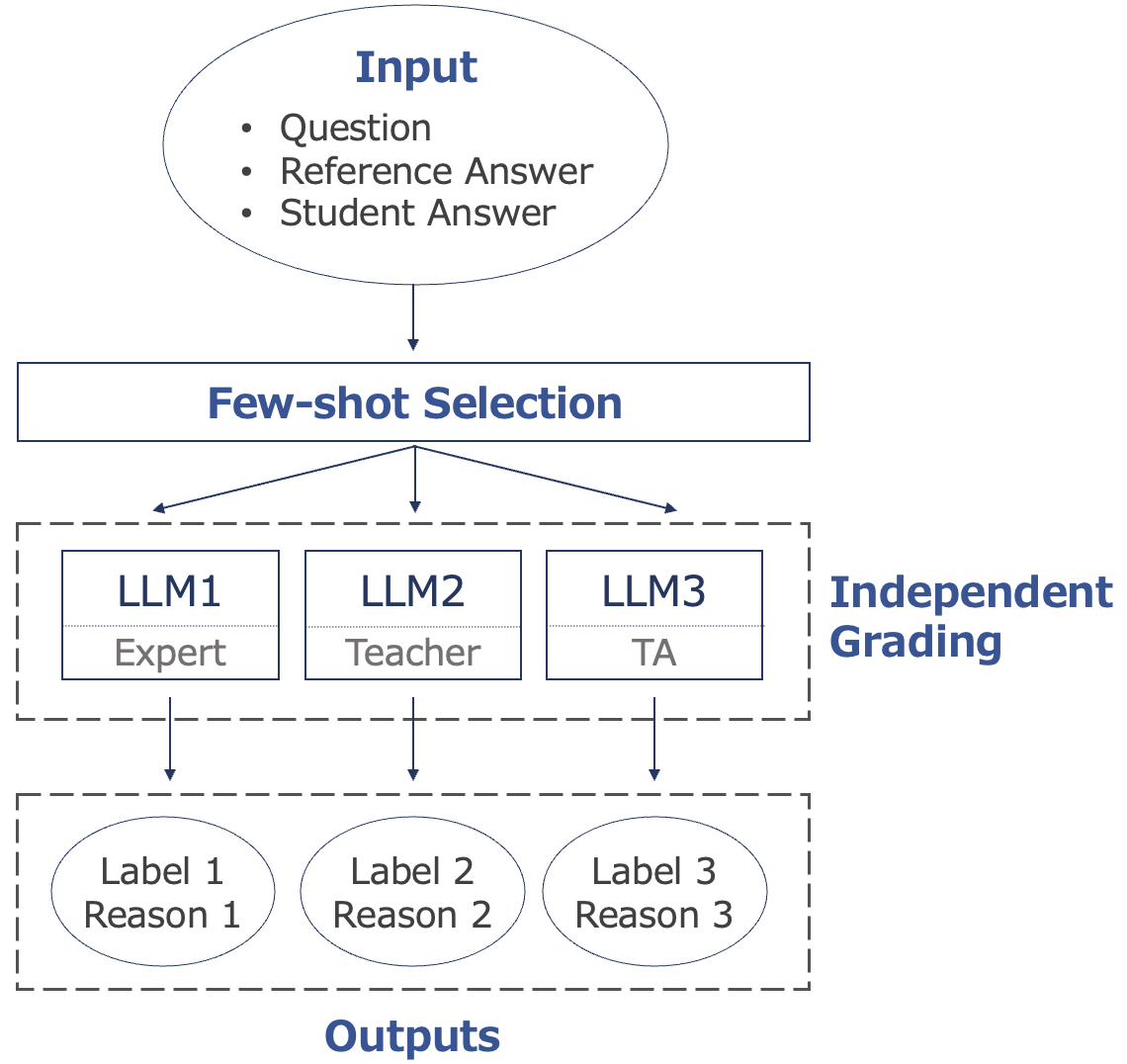}
        \caption{\textbf{Process Flow of Multi-LLM Grading}}
        \label{fig-multillmgrading}
    \end{minipage}

    \vspace{0.5cm}

    \begin{minipage}[b]{0.55\textwidth}
        \centering
        \includegraphics[width=\textwidth]{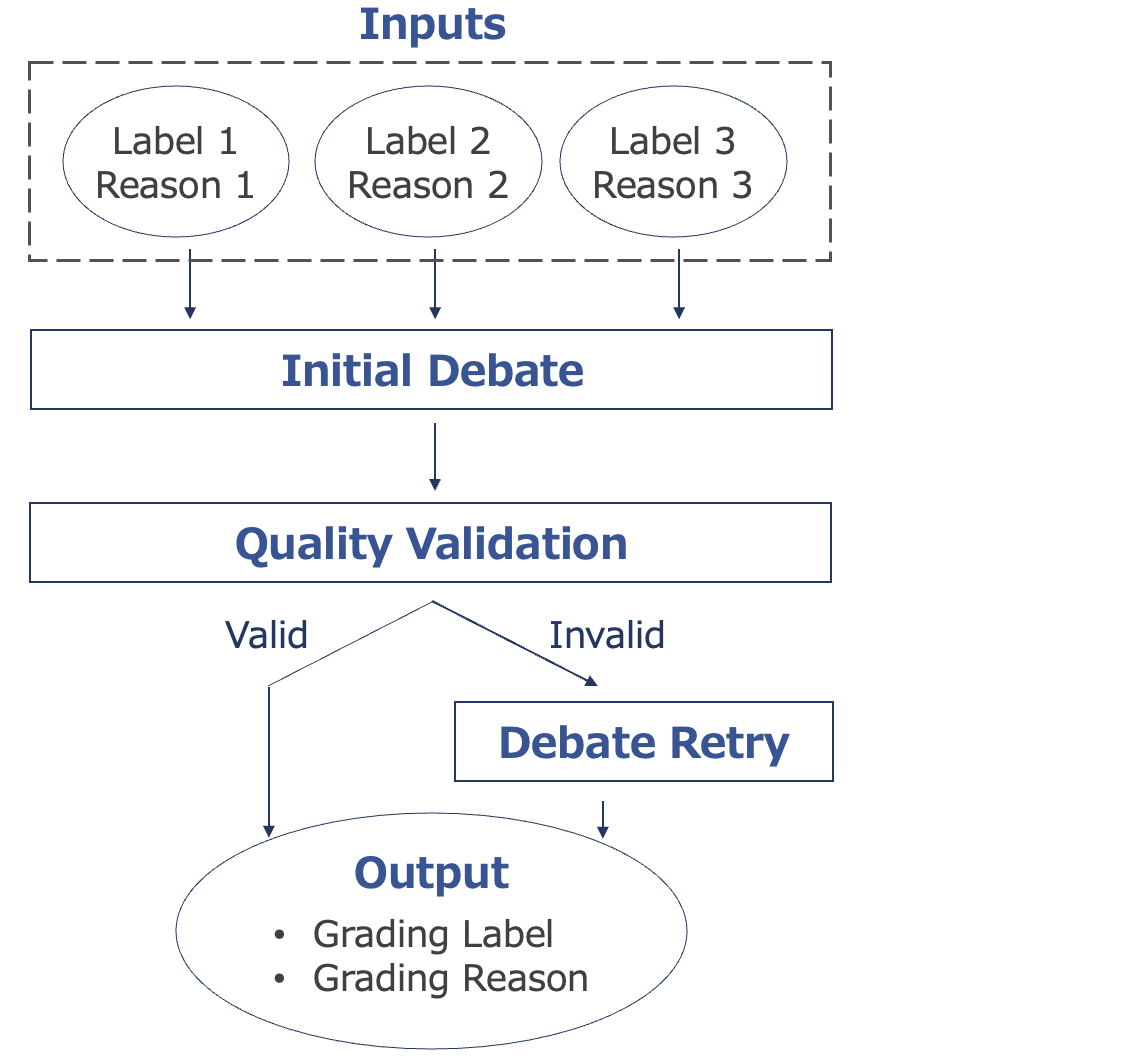}
        \caption{\textbf{Process Flow of Debate Integration}}
        \label{fig-debateflow}
    \end{minipage}
\end{figure}

\textit{Multi-LLM grading} involves two steps: \textit{few-shot selection} and \textit{independent grading}, as shown in Fig.\,\ref{fig-multillmgrading}.
First, \textit{few-shot selection} chooses three grading examples from the dataset $D'$.
In \textit{independent grading}, three LLMs separately grade the student answers based on these examples.
Finally, the grading results are forwarded to the \textit{debate integration} phase (illustrated in Fig.\,\ref{fig-debateflow} and detailed in Section \ref{subsec-debateintegration}).

\subsubsection{Few-shot Selection}
We employ a three-shot prompt (instructions provided to LLMs with three output examples) for the grading task.
This approach is motivated by the findings of Fateen et al. \citep{beyondscores}, who demonstrated that selecting appropriate examples for each student answer using ColBERT \citep{colbert} significantly enhances grading performance.
Based on their study, the GET system incorporates a retrieval phase to search for suitable grading cases.
This process involves the following steps:

\begin{enumerate}
    \item \textbf{Retrieve relevant grading examples:}
          \begin{itemize}
              \item If the question $q_i$ has been posed to students in $D'$, retrieve the corresponding grading examples: ($q_i$, $r_i$, $s'_{i_l}$, $gl'_{i_l}$, $gr'_{i_l}$).
              \item Otherwise, gets all data ($q'_k$, $r'_k$, $s'_{k_l}$, $gl'_{k_l}$, $gr'_{k_l}$) $\in D'$.
          \end{itemize}

    \item \textbf{Prepare data for embedding:}
          Concatenate all items in each grading example into a single string.

    \item \textbf{Generate embeddings:}
          \begin{itemize}
              \item Generate token embeddings (vector representations capturing semantic meaning) for the concatenated examples by ColBERT's document encoder.
              \item Generate token embeddings for the student answer $s_{i_j}$ by ColBERT's query encoder.
          \end{itemize}

    \item \textbf{Compute cosine similarities:}
          \begin{itemize}
              \item For each token in $s_{i_j}$, calculate its cosine similarity with the tokens in the grading examples.
              \item Identify the maximum similarity value for each token in $s_{i_j}$.
          \end{itemize}

    \item \textbf{Aggregate similarity scores:}
          Sum the maximum similarity scores across all tokens in $s_{i_j}$.

    \item \textbf{Rank and select examples:}
          Rank the grading examples based on their total similarity scores in descending order. Select the top three examples for grading.
\end{enumerate}

The selected examples are reformatted to guide LLM outputs when grading.
In this format, the grading label is presented first, followed by the reason.
This sequence is adopted because Jiang et al. \citep{sasgpt4} found that placing the label before the reason improves accuracy compared to reason-first approaches.

\subsubsection{Independent Grading}
After selecting examples, three different LLMs independently grade the student answer $s_{i_j}$ by providing both a grading label and reason.

\begin{figure}[!b]
    \centering
    \includegraphics[width=\textwidth]{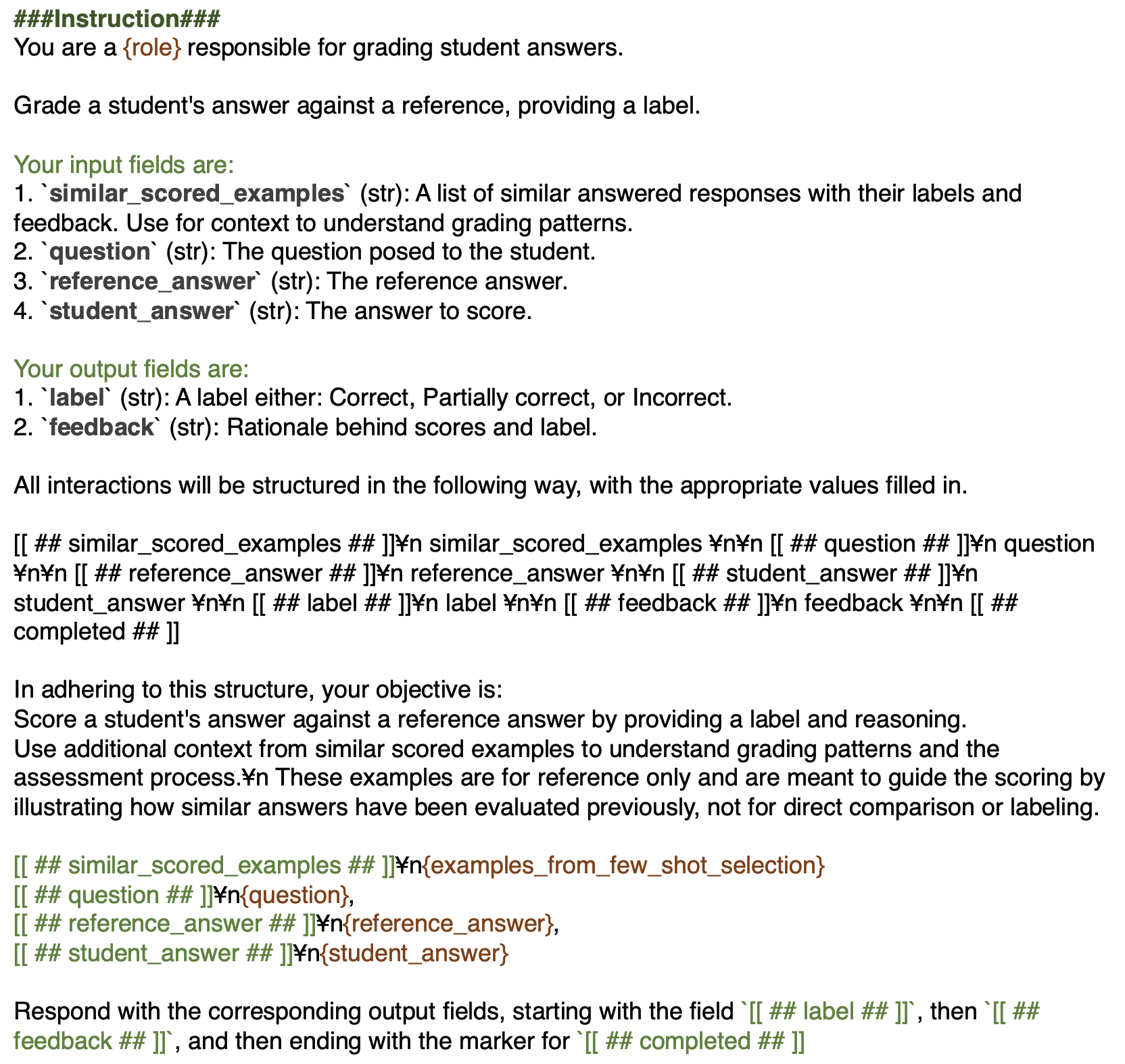}
    \caption{\textbf{Example prompt for grading student answers.} The prompt outlines the grading task and includes three specific grading examples selected in the previous step. For questions without past data, supplemental information is provided to guide grading (shown in Fig.\,\ref{fig-detcri}).}
    \label{fig-singlegrading}
\end{figure}

\begin{figure}[!tb]
    \centering
    \includegraphics[width=0.9\textwidth]{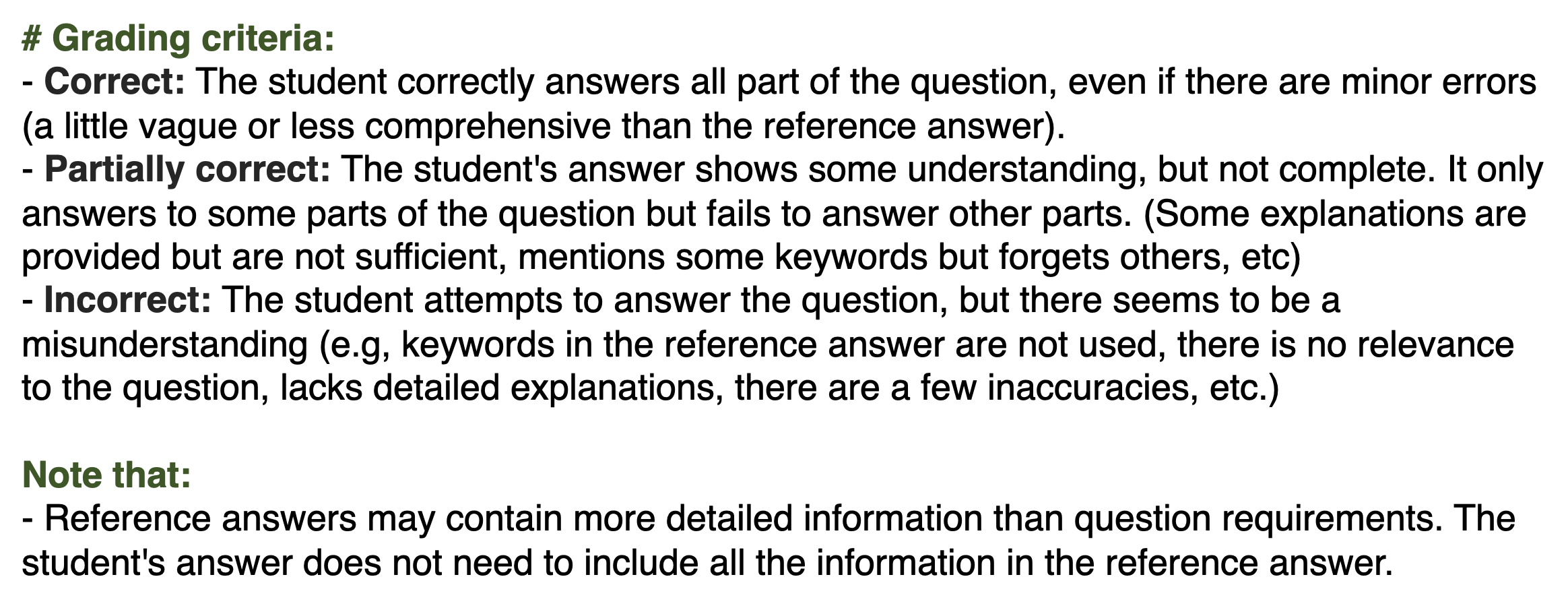}
    \caption{\textbf{Additional information to guide grading without the same question in the past data} This supplements LLMs by explaining grading policy.}
    \label{fig-detcri}
\end{figure}

The grading prompt is illustrated in Fig.\,\ref{fig-singlegrading} and Fig.\,\ref{fig-detcri}.
We created this prompt by referring to the GitHub code published by Fateen et al \citep{beyondscores}.
The prompt is almost similar to that used in the \textit{grading past-data} of the \textit{pseudo-learning} phase.
However, in \textit{independent grading}, we additionally assign distinct role names to the LLMs to enhance the diversity of output.
According to a previous study, combining models with different characteristics improves the performance more than combining similar models \citep{ensembleoflm}.
Inspired by this, we assign different roles for LLMs based on their macro F1 scores in \textit{pseudo-learning}.
The model with the highest score is assigned ``Skilled Expert Grader'', the second-highest is ``University Teacher'', and the lowest is ``Student TA''.

\subsection{Debate Integration}\label{subsec-debateintegration}

After \textit{multi-LLM grading} by three LLMs, one of them integrates the results into a unified grading conclusion.
For this task, the LLM with the highest macro F1 score in the \textit{pseudo-learning} phase, designated as the Skilled Expert Grader, is utilized.

The \textit{debate integration} process has three steps: \textit{initial debate}, \textit{quality validation}, and \textit{debate retry}, as shown in Fig.\,\ref{fig-debateflow}

\subsubsection{Initial Debate}
\begin{figure}[!htb] \centering \includegraphics[width=0.95\textwidth]{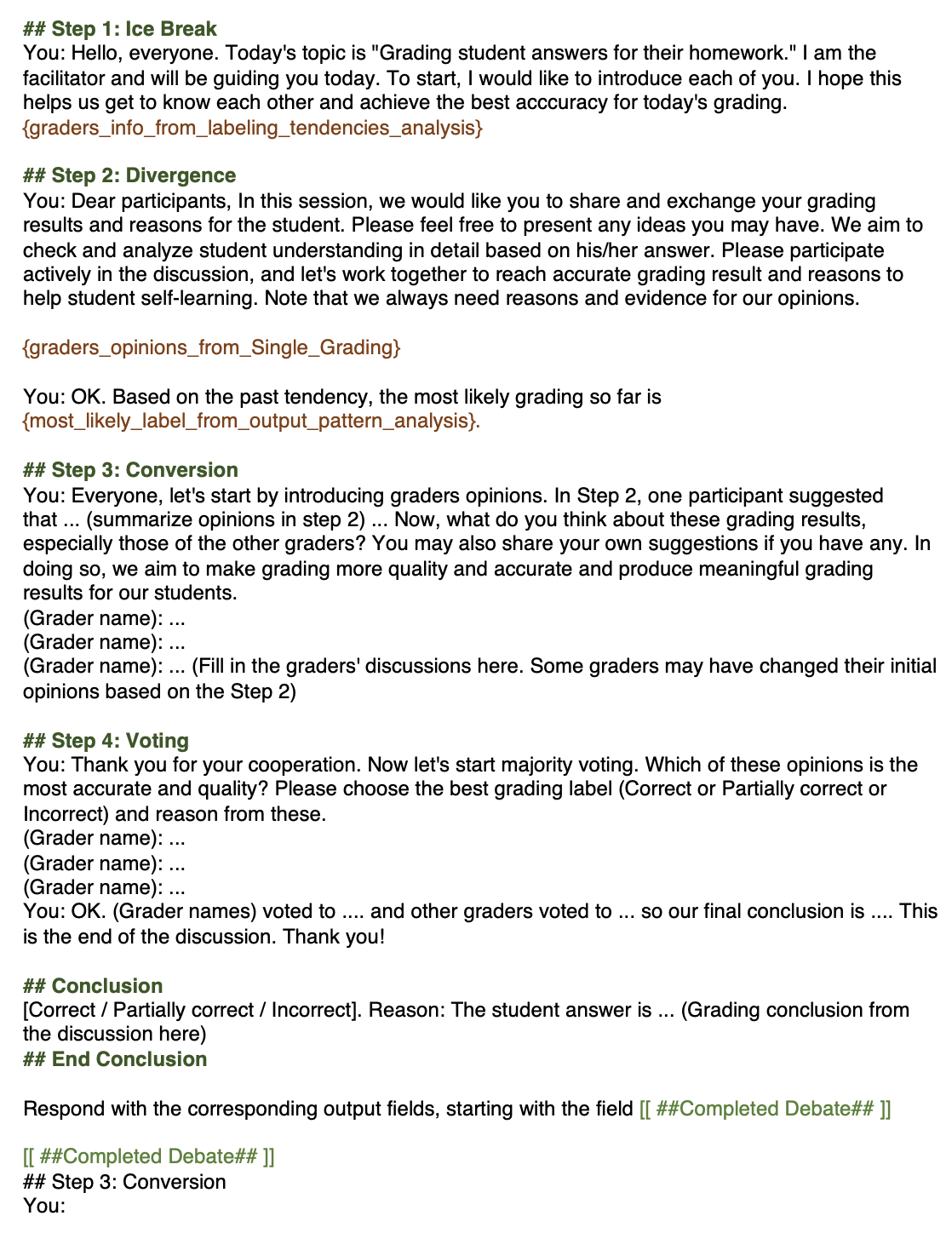}
    \caption{\textbf{Debate template.} This template is provided to the LLM for completion during the debate integration process.}
    \label{fig-debateintegration}
\end{figure}

In the \textit{initial debate}, the selected LLM  simulates a debate among the graders by completing a template as shown in Fig.\,\ref{fig-debateintegration}.
The debate follows four stages: Ice Break, Divergence, Conversion, and Voting.
This debate construction is inspired by the work of Dong et al. who propose an effective way to facilitate debates using LLMs \citep{llmfacilitator}.

The statements made by the three graders during the debate are not the actual outputs of three separate LLMs.
Instead, they are generated by a single LLM.
This single LLM generates the statements based on the grading results produced during the \textit{multi-LLM grading} stage and the grading tendencies identified in the \textit{pseudo-learning} phase.

Additionally, if the question $q_i$ exists in the past dataset $D'$, the three examples selected during the \textit{few-shot selection} stage are also included as the input.
Conversely, if $q_i$ is not present in $D'$, no examples are provided.
This approach prevents the LLM from over-relying on past data when grading new questions, thereby ensuring high-quality performance across diverse datasets.

\subsubsection{Quality Validation}
The \textit{quality validation} phase ensures the quality and correctness of the results integrated in the previous step.
The same LLM used in the debate is employed for this validation.

As the input, it takes $q_i$, $r_i$, $s_{i_j}$, the most likely label for $s_{i_j}$ (same as passed in the \textit{debate integration} phase), and the integrated grading label and reason to be evaluated.
Additionally, if the question $q_i$ exists in the historical dataset $D'$, the grading result generated by it during \textit{independent grading} is also provided as a grading example.
Conversely, if $q_i$ is not found in $D'$, no examples are given.
This is because the grading performance of each LLM can vary significantly depending on the question.
Notably, the best-performing LLM in \textit{pseudo-learning} does not necessarily excel when grading a novel question $q_i$.

If the result is deemed valid, it is finalized as the grading conclusion.
Otherwise, the LLM outputs the revised grading result.
Based on this, the following \textit{debate retry} phase determines the final grading conclusion.

Note that the \textit{quality validation} phase is skipped if the grading labels produced by the three LLMs during \textit{independent grading} and the final label from the discussion are all identical.
Empirical observations from the experiment indicate that when all LLMs agree, and the discussion result aligns with this consensus, further validation does not alter the conclusion.

\subsubsection{Debate Retry}
\begin{figure}[!htb]
    \centering
    \includegraphics[width=\textwidth]{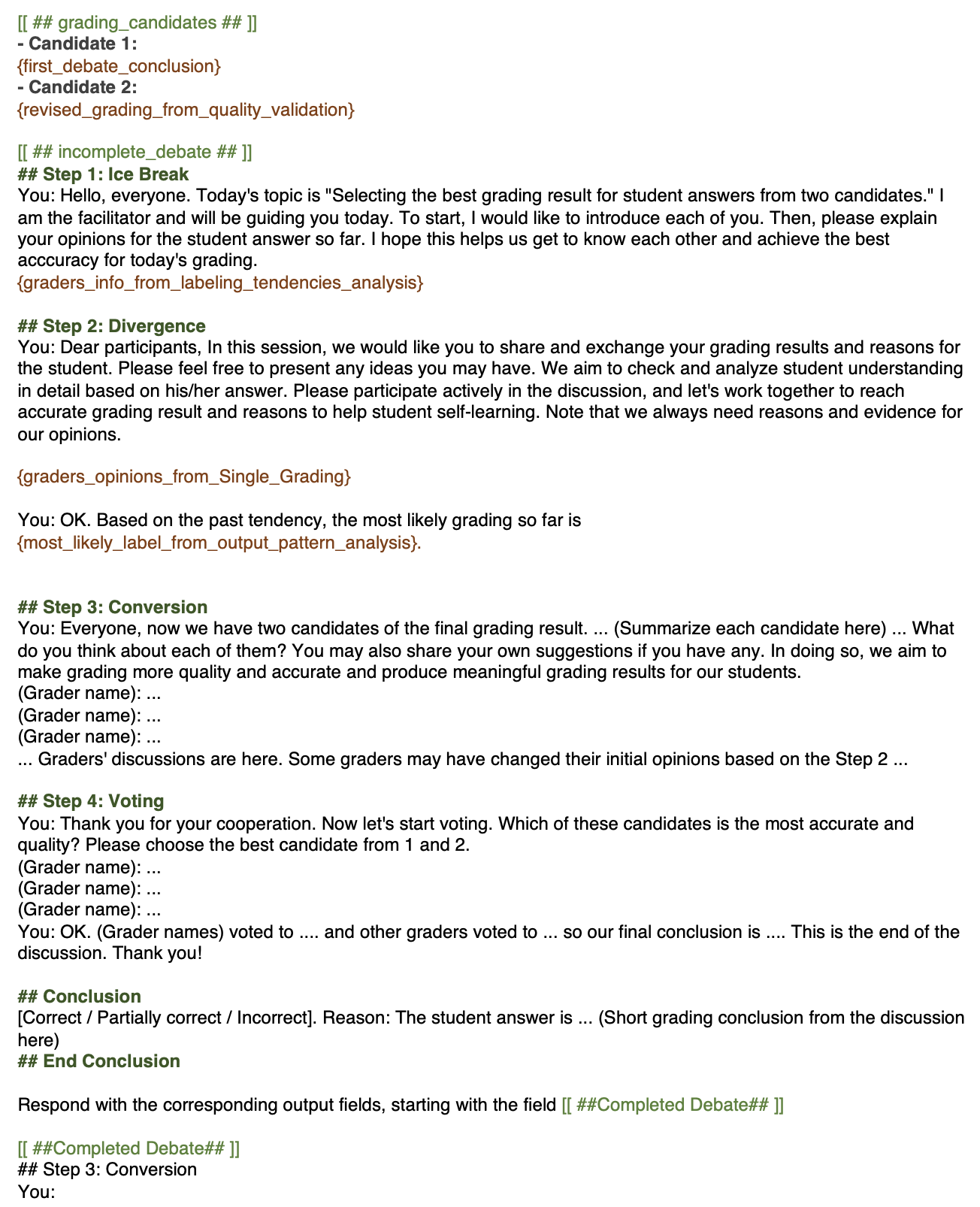}
    \caption{\textbf{Debate retry template}: The prompt instructs the LLM to compare two grading candidates via a discussion. }
    \label{fig-debateretrt}
\end{figure}

The \textit{debate retry} phase is executed only if the integrated grading result is deemed invalid in the \textit{quality validation} phase.
In this phase, the LLM compares the first debate result and revised grading from the \textit{quality validation}.
This process is also done by simulating debate among three graders.
The debate template for this is shown in Fig.\,\ref{fig-debateretrt}.
The output from this phase is determined as the final conclusion.

\section{Experimental Setup}\label{sec-experriments}
We evaluated our system's grading performance through automated assessments.
These assessments measured its ability to predict grading labels and provide valid reasoning.
This section outlines the experimental settings.

\subsection{Dataset}
We used the Short Answer Feedback (SAF) Dataset \citep{safdataset} for evaluation.
This dataset targets a college-level communication network course and contains questions, reference answers, student answers, grading labels (Correct, Partially correct, Incorrect), and grader feedback.

It is divided into four subsets: train, validation, test-unseen-answers (UA), and test-unseen-questions (UQ).
The subsets contain 1700, 427, 375, and 479 entries, respectively.
The train set and UA set contain the same questions but include different student answers.
In contrast, the train set and UQ set have neither overlapping questions nor student answers.
We used the train set for \textit{pseudo-learning} and the UA and UQ sets to evaluate system performance.

\subsection{Evaluation Methods}
\subsubsection{Statistic Evaluation}
We evaluated grading label and reason performance using statistical metrics, following previous studies.

Since grading label prediction is a task of classifying student answers into three categories (Correct, Partially correct, and Incorrect), accuracy and macro F1 score were used as evaluation metrics.
The definitions of these metrics are shown in Eq.\,\ref{eq-acc} and Eq.\,\ref{eq-macrof1}.

For the generation of grading reasons, the grader feedbacks included in the SAF dataset were used as a reference, and the results were evaluated using BLEU \citep{bleu}, ROUGE-2 \citep{rouge}, and BERTScore \citep{bertscore}.
However, BLEU and ROUGE measure word similarity rather than semantic similarity, which may not fully reflect the quality of the grading reasons \citep{beyondscores}.
BERTScore also uses word embeddings, so it tends to be higher when the LLM reproduces the same phrases as the dataset's grading reasons.
Therefore, these metrics are shown for reference only, and the quality of the grading reasons will be mainly evaluated using LLMs, as described below.

\subsubsection{LLM-Based Evaluation}
To address the limitations of statistical evaluation for grading reasons, we introduced an automated evaluation process using LLMs.
This approach is inspired by Jurenka et al. \citep{towardsresponsible}, who utilized LLMs to validate the quality of hints provided to students.
Building on this concept, we designed a new process to assess the validity of the grading reasons generated by our system and baselines.

The evaluation is conducted as follows:
\begin{enumerate}
    \item Each of the three LLMs used in the GET system individually assesses the validity of the grading reasons.
    \item The percentage of grading reasons rated as valid by each LLM is calculated.
    \item The final evaluation result is obtained by averaging these percentages.
\end{enumerate}

\begin{figure}[!htb]
    \centering
    \includegraphics[width=0.85\textwidth]{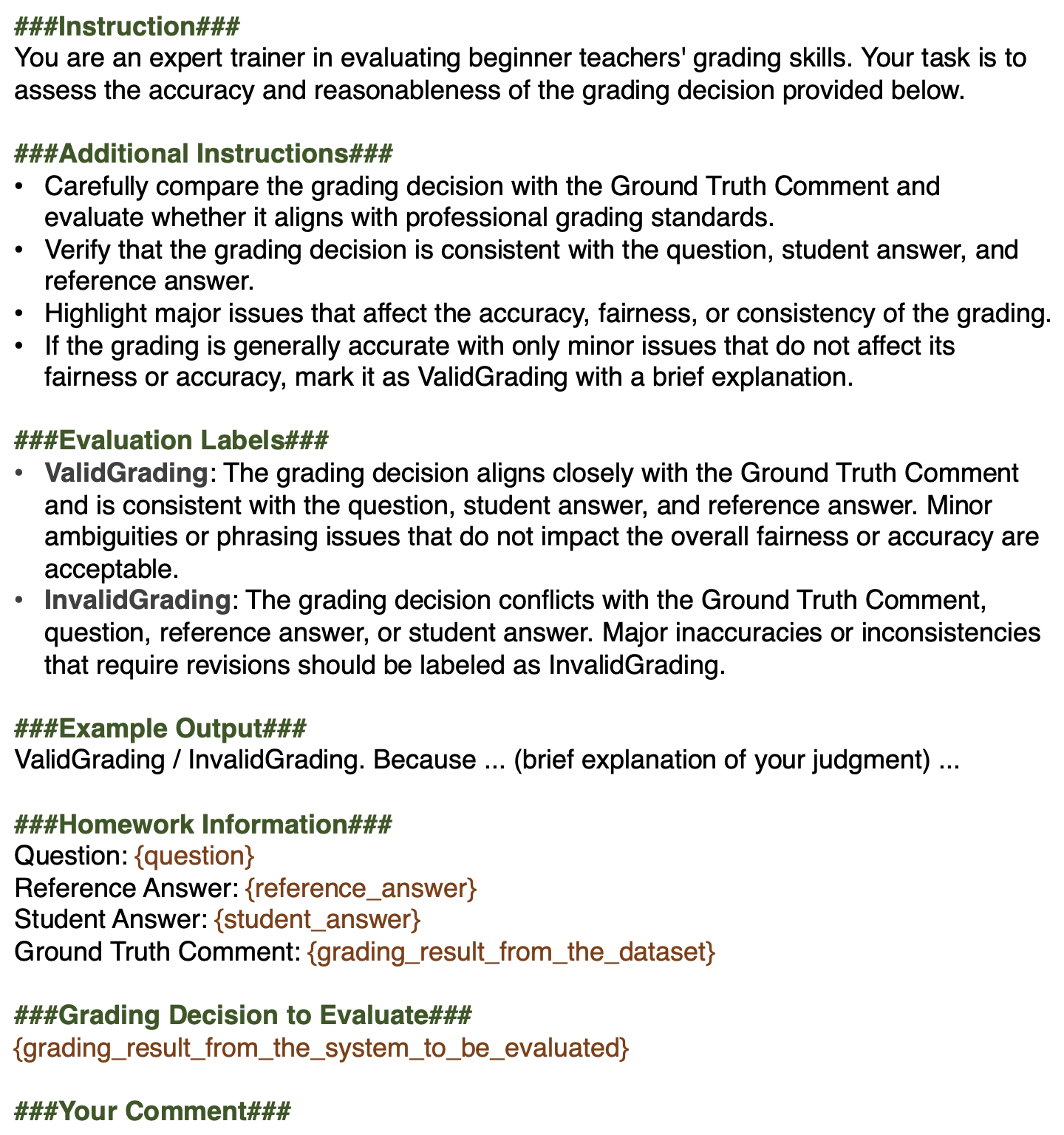}
    \caption{\textbf{The Prompt used in LLM-Based Evaluation}}
    \label{fig-llmeval}
\end{figure}

The evaluation prompt is shown in Fig.\,\ref{fig-llmeval}.
It includes the grading reasons recorded in the datasets alongside the system-generated grading results.
The LLM compares these inputs to determine the validity of the grading reasons.

Unlike the approach in \citep{towardsresponsible}, which relies on a single LLM for evaluation, our method aggregates results from three LLMs.
By averaging their evaluations, we aim to achieve a more objective and balanced assessment, avoiding reliance on a specific LLM.

\subsection{Model Selection and Parameter Settings}
To evaluate the GET system, we selected LLMs based on usability and cost-effectiveness to ensure practical implementation in real educational environments.

\subsubsection{Models in Pseudo-Learning}
To describe label tendencies using classification metrics, we employed Gemini 1.5 Flash \citep{gemini}.
Among the three models used in \textit{multi-LLM grading} (introduced in the following section), Gemini 1.5 Flash was chosen due to its superior performance on mathematics-related tasks, as reported in benchmarks \citep{llamabench}, \citep{geminibenck}, \citep{mixtralbench} (Accessed: December 26, 2024).
Since the task involves analyzing numerical metrics such as accuracy and F1 scores, Gemini's capabilities align closely with the requirements of this analysis.

\subsubsection{Models in Multi-LLM Grading}
For grading tasks, we selected Meta-Llama-3-8B-Instruct \citep{llama3modelcard}, Gemini 1.5 Flash \citep{gemini}, and Mixtral-8x22b \citep{mixtral}.
These models were chosen because they either offer free-tier APIs or are open-source, making them cost-effective and accessible for integration into the GET system.

\subsubsection{Parameters}
In \textit{pseudo-learning}, the output pattern analysis (Algorithm \ref{alg-mostlikelylabel}) needs the threshold to decide the most likely label for each output combination.
In the experiment, it is set to 1.2 based on the results of trials with several values.

Three LLMs also have some parameters.
We set the temperature of them to 0.7.
The number of maximum tokens which newly generated is set to 8192, which is the same as the maximum length of ColBERT's input.
Other parameters of each LLM are left as default.

\subsection{Pseudo-learning on the selected LLMs}
Before the main experiments, we conducted a \textit{pseudo-learning} phase.
The results of \textit{grading past-data} are summarized in Table \ref{tab-metricsscores}.
\begin{table}[!tb]
    \centering
    \caption{Metrics Scores for Models}
    \label{tab-metricsscores}
    \renewcommand{\arraystretch}{1.3}
    \begin{tabular}{lccc}
        \hline
        \textbf{Metric}              & \textbf{Llama-3-8B-it} & \textbf{Gemini-1.5-Flash} & \textbf{Mixtral-8x22B} \\ \hline
        Accuracy                     & 0.5153                 & 0.5206                    & 0.6282                 \\
        F1 Score (Correct)           & 0.5385                 & 0.4709                    & 0.7066                 \\
        F1 Score (Partially Correct) & 0.5118                 & 0.5594                    & 0.5290                 \\
        F1 Score (Incorrect)         & 0.3492                 & 0.5205                    & 0.4818                 \\
        \textbf{Macro F1}            & 0.4665                 & 0.5170                    & \textbf{0.5725}        \\ \hline
    \end{tabular}%
\end{table}

Based on their macro F1 scores, role names were assigned to each LLM: Mixtral-8x22b was designated as the Skilled Expert Grader, Gemini 1.5 Flash as the University Teacher, and Llama-3-8B-Instruct as the Student TA.

\begin{figure}[!htb]
    \centering
    \includegraphics[width=0.85\textwidth]{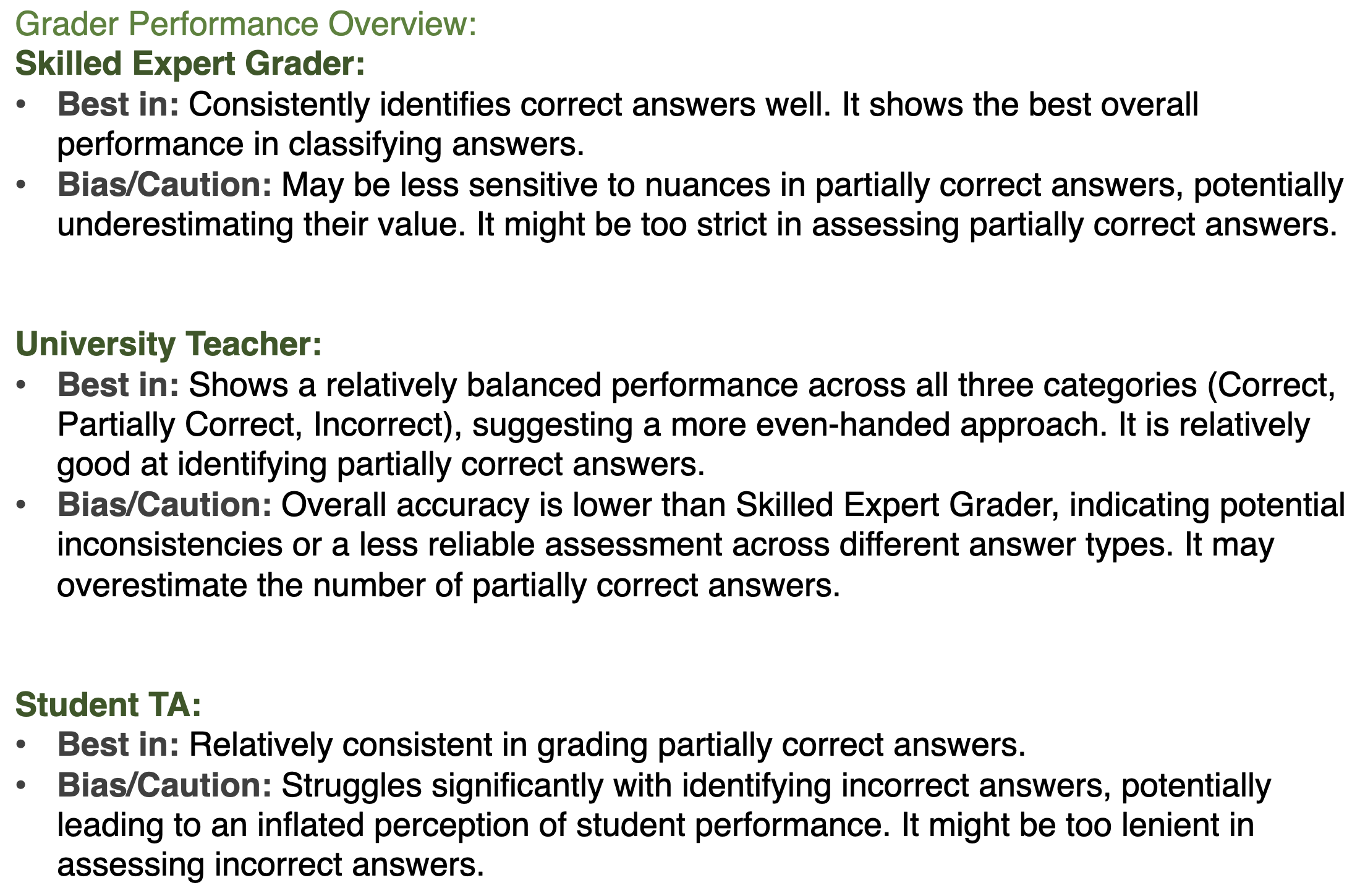}
    \caption{\textbf{The result of Label tendencies analysis}}
    \label{fig-labeltendres}
\end{figure}

The results of the \textit{tendencies identification} phase are presented in Fig.\,\ref{fig-labeltendres} (Labeling tendencies analysis) and Table \ref{tab-mostlikelylabelres} (Output pattern analysis).

\begin{table}[!htb]
    \centering
    \caption{Summary of model predictions and most likely labels}
    \label{tab-mostlikelylabelres}
    \renewcommand{\arraystretch}{1.3}
    \begin{tabular}{lllc}
        \hline
        \textbf{Llama-3-8b-it} & \textbf{Gemini-1.5-Flash} & \textbf{Mixtral-8x22b} & \textbf{Most Likely Label}                                         \\ \hline
        Correct                & Correct                   & Correct                & Correct                                                            \\
        Correct                & Partially correct         & Correct                & Correct                                                            \\
        Partially correct      & Correct                   & Correct                & Correct                                                            \\
        Partially correct      & Correct                   & Partially correct      & Correct                                                            \\
        Partially correct      & Partially correct         & Correct                & Correct                                                            \\ Incorrect & Correct & Correct & Correct                                                \\\hline
        Partially correct      & Partially correct         & Partially correct      & Partially correct                                                  \\
        Correct                & Partially correct         & Partially correct      & Partially correct                                                  \\
        Incorrect              & Partially correct         & Correct                & Partially correct                                                  \\
        Correct                & Partially correct         & Incorrect              & Partially correct                                                  \\  \hline
        Incorrect              & Partially correct         & Partially correct      & \begin{tabular}{c} Incorrect or \\ Partially correct \end{tabular} \\
        Partially correct      & Incorrect                 & Partially correct      & \begin{tabular}{c} Incorrect or \\ Partially correct \end{tabular} \\
        Incorrect              & Partially correct         & Incorrect              & \begin{tabular}{c} Incorrect or \\ Partially correct \end{tabular} \\ \hline
        Correct                & Correct                   & Partially correct      & \begin{tabular}{c} Incorrect or \\ Correct \end{tabular}           \\ \hline
        Incorrect              & Incorrect                 & Incorrect              & Incorrect                                                          \\
        Incorrect              & Incorrect                 & Partially correct      & Incorrect                                                          \\
        Partially correct      & Incorrect                 & Incorrect              & Incorrect                                                          \\
        Partially correct      & Partially correct         & Incorrect              & Incorrect                                                          \\ \hline
    \end{tabular}
\end{table}

\subsection{Baseline Models}
To evaluate the grading performance of the GET system, we compared it against several baseline systems.
These baselines fall into two main categories: grading methods for \textbf{comparison study} and \textbf{ablation study}.

\subsubsection{Grading Methods for Comparison Study}
These methods provide a benchmark to compare the performance of our proposed system against existing approaches:
\begin{itemize}
    \item \textbf{Majority Voting Baseline}: Grading labels are determined based on the majority vote among three graders. The grading reason is randomly selected from the graders who align with the majority label. This follows the same flow of \textit{multi-LLM grading} to the GET system, but is different in using the majority voting instead of \textit{debate integration}.
    \item \textbf{Criteria-Based Grading System} \citep{icetc}: A system we introduced at ICETC 2024. It decomposes each question into specific grading criteria based on predefined grammatical rules. A fine-tuned language model (Zephyr-7B-Beta) then evaluates the student answers based on the criteria to generate grading labels and their reasons.
    \item \textbf{ASAS-F-RAG} \citep{beyondscores}: A system by Fateen et al., which is constructed from an LLM with RAG. We implemented this approach using their publicly available GitHub repository for replication.
\end{itemize}

\subsubsection{Grading Methods for Ablation Study}
These variants are designed for an ablation study, helping us assess the impact of individual components in the GET system:
\begin{enumerate}
    \item \textbf{Single LLM Graders}: Each LLM (Mixtral-8x22b, Llama-3-8B-Instruct, and Gemini 1.5 Flash) independently grades student answers without the Ensemble ToT framework. Grading is performed through the same steps of \textit{multi-LLM grading} phase.
    \item \textbf{Proposed System Without Pseudo-Learning}: This variant skips the pseudo-learning phase and only performs \textit{multi-LLM grading} and \textit{debate integration}. Without \textit{pseudo-learning}, the system cannot leverage the performance tendencies of each grader. Furthermore, all LLMs are assigned the same generic role name (\textit{Grader}) rather than specific roles derived from \textit{pseudo-learning} results.
    \item \textbf{Proposed System with a Single LLM}: In this variant, the GET system uses a single LLM (Mixtral-8x22b) instead of three different LLMs. Since \textit{pseudo-learning} depends on comparing outputs from multiple LLMs, it is omitted. To simulate \textit{multi-LLM grading}, the same LLM grades each student answer three times, with different role names assigned for each instance.
\end{enumerate}

These baseline models allow us to evaluate the robustness of the GET system and analyze the effectiveness of its components.

\section{Experimental Results and Discussion}
We evaluated the proposed system and baseline methods on two subsets of the SAF dataset: UA and UQ.
First, we present the comparison results between the GET system and other grading methods.
This is followed by the results of an ablation study using partial variants of GET.

\subsection{Comparison Study}
This subsection presents the results of the GET system compared to other grading methods.

\subsubsection{Grading Label Prediction Analysis}
The results of the statistical evaluation of grading label predictions are presented in Tables \ref{table-compare_acc_f1}.
All indicators are rounded to the fifth decimal place.

\begin{table}[!b]
    \centering
    \caption{Accuracy and F1 Score of GET and other grading systems}
    \label{table-compare_acc_f1}
    \renewcommand{\arraystretch}{1.3}
    \begin{tabular}{lcccccc}
        \hline
        \textbf{}      & \multicolumn{3}{c}{\textbf{Accuracy}} & \multicolumn{3}{c}{\textbf{F1 Score}}                                                                         \\\hline
                       & UA                                    & UQ                                    & Avg.            & UA              & UQ              & Avg.            \\ \hline
        Majority Vote  & 0.7760                                & 0.6388                                & 0.7074          & 0.7113          & 0.6111          & 0.6621          \\
        Criteria-Based & 0.7387                                & 0.5595                                & 0.6491          & 0.5496          & 0.5525          & 0.5511          \\
        ASAS-F-RAG     & 0.7093                                & 0.6534                                & 0.6814          & 0.6315          & \textbf{0.6289} & 0.6302          \\
        \textbf{GET}   & \textbf{0.7787}                       & \textbf{0.6701}                       & \textbf{0.7244} & \textbf{0.7128} & 0.6268          & \textbf{0.6698} \\ \hline
    \end{tabular}
\end{table}

The proposed system achieved the highest macro F1 score and accuracy on the UA dataset, as well as the highest accuracy on the UQ dataset.
These results emphasize the effectiveness of our approach, which combines outputs from multiple models.
Furthermore, the system demonstrated consistent performance across different dataset types, as reflected in its highest average accuracy and F1 score among all baselines.

\subsubsection{Grading Reason Generation Analysis}
The results of the LLM-based evaluation are summarized in Table \ref{table-compare_llmcritic}.

\begin{table}[!tb]
    \centering
    \caption{LLM-Based Evaluation results of GET and other grading systems}
    \label{table-compare_llmcritic}
    \renewcommand{\arraystretch}{1.3}
    \begin{tabular}{lccc}
        \hline \textbf{} & UA              & UQ              & Avg.            \\ \hline
        Majority Voting  & \textbf{0.7040} & 0.6326          & 0.6683          \\ 
        Criteria-Based   & 0.4267          & 0.1635          & 0.2951          \\ 
        ASAS-F-RAG       & 0.6355          & 0.6145          & 0.6250          \\ 
        \textbf{GET}     & 0.7022          & \textbf{0.6360} & \textbf{0.6691} \\ \hline 
    \end{tabular}
\end{table}

The proposed system achieved the highest average proportion of valid grading reasons for the UQ dataset and in the combined average across both datasets.
These results indicate that the GET system generates semantically accurate grading reasons more effectively than baseline systems.
This supports the effectiveness of the Ensemble ToT framework for both label prediction (classification) and grading reason generation (text generation).

However, on the UA dataset, the majority vote baseline outperformed the proposed system.
This outcome appears to stem from the overly concise summarization of grading reasons during the debate phase, which sometimes resulted in a loss of detail.
For instance, consider the following outputs for the same student answer in the UA set:

\begin{itemize}
    \item \textbf{A grading reason from the majority vote baseline:} \\
          \textit{Incorrect. The student's answer incorrectly states that the system will never reach equilibrium. While the arrival rate is less than the service rate, this doesn't prevent the system from reaching a steady state. The question asks for the expected time the system will have fewer than 10 packets, which requires calculating the probability of the system being full and subtracting that from the total observation time. The student did not address this core aspect of the problem. Similar examples demonstrate that a correct answer involves calculating the blocking probability and using it to determine the expected time the system is not full.}

    \item \textbf{A grading reason from the GET system:} \\
          \textit{Incorrect. Reason: The student's answer is incorrect because it misunderstands the concept of equilibrium in a queuing system and lacks a clear calculation for the expected time the system is not blocked.}
\end{itemize}

In this example, the majority vote baseline, which directly uses outputs from multiple LLMs without debate, provided a more detailed and contextually rich response.
While GET's output is not inaccurate, it lacks the depth and specific details present in the baseline's response.
This lack of detail was noted by the LLM evaluator. This over-summarization contributed to the baseline's superior performance on the UA dataset.

Conversely, GET achieved higher scores on the UQ dataset.
This is likely because the majority vote baseline's performance on UQ is hampered by its inability to access past grading examples for similar questions, negatively impacting output quality.
GET, however, mitigates this limitation through its Ensemble ToT-based architecture, incorporating \textit{pseudo-learning} and \textit{debate integration} to maintain output quality.

Finally, for reference, the results of the statistical evaluation of the quality of the grading reasons are presented in Table \ref{table-compare_bleu_rouge_bertscore}.
While the GET system's higher performance in LLM-based evaluation, it did not achieve the best performance in BLEU, ROUGE, or BERTScore.
For these metrics, the criteria-based grading system achieved superior scores.
This discrepancy arises from the criteria-based system being fine-tuned on the SAF dataset's train set, enabling it to produce outputs with phrasing more closely aligned with the reference text.

\begin{table}[!b]
    \centering
    \caption{BLEU, ROUGE, and BERTScore of GET and other grading systems}
    \label{table-compare_bleu_rouge_bertscore}
    \renewcommand{\arraystretch}{1.3}
    \begin{tabular}{lcccccc}
        \hline
        \textbf{}      & \multicolumn{2}{c}{\textbf{BLEU}} & \multicolumn{2}{c}{\textbf{ROUGE}} & \multicolumn{2}{c}{\textbf{BERTScore}}                                                       \\\hline
                       & UA                                & UQ                                 & UA                                     & UQ              & UA              & UQ              \\ \hline
        Majority Vote  & 0.0392                            & 0.0234                             & 0.0896                                 & 0.0700          & 0.7005          & 0.7044          \\
        Criteria-Based & \textbf{0.1559}                   & \textbf{0.0548}                    & \textbf{0.2573}                        & \textbf{0.1349} & \textbf{0.7218} & 0.6847          \\
        ASAS-F-RAG     & 0.1024                            & 0.0337                             & 0.1647                                 & 0.0972          & 0.7212          & 0.7132          \\
        \textbf{GET}   & 0.0288                            & 0.0289                             & 0.0682                                 & 0.0772          & 0.6953          & \textbf{0.7171} \\ \hline
    \end{tabular}
\end{table}

\subsection{Ablation Study}
This subsection discusses the results of the ablation study with partial variants of the GET system.
\subsubsection{Grading Label Prediction Analysis}
The evaluation results for grading label prediction are presented in Tables \ref{table-ablation_acc_f1}.

\begin{table}[!b]
    \centering
    \caption{Accuracy and F1 Score of GET and its partial variants}
    \label{table-ablation_acc_f1}
    \renewcommand{\arraystretch}{1.3}
    \begin{tabular}{lcccccc}
        \hline
        \textbf{}        & \multicolumn{3}{c}{\textbf{Accuracy}} & \multicolumn{3}{c}{\textbf{F1 Score}}                                                                         \\ \hline
                         & UA                                    & UQ                                    & Avg.            & UA              & UQ              & Avg.            \\ \hline
        Mixtral-8x22b    & 0.7680                                & 0.6284                                & 0.6982          & 0.6686          & 0.5820          & 0.6253          \\
        Gemini 1.5 Flash & 0.7387                                & 0.6472                                & 0.6929          & 0.7110          & 0.6146          & 0.6628          \\
        Llama3-8b-it     & 0.6987                                & 0.4760                                & 0.5873          & 0.5809          & 0.4622          & 0.5215          \\
        W/O Pseudo       & \textbf{0.7787}                       & 0.65762                               & 0.7181          & \textbf{0.7288} & 0.6185          & \textbf{0.6737} \\
        W Single LLM     & 0.7600                                & 0.6326                                & 0.6963          & 0.6539          & 0.5830          & 0.6184          \\
        \textbf{GET}     & \textbf{0.7787}                       & \textbf{0.6701}                       & \textbf{0.7244} & 0.7128          & \textbf{0.6268} & 0.6698          \\ \hline
    \end{tabular}
\end{table}

The proposed system outperformed all baselines in terms of accuracy on the UA and UQ sets.
In terms of the F1-score, it achieves the highest in the UQ set.

The system outperformed all single-LLM baselines, demonstrating the effectiveness of the Ensemble ToT framework.
Additionally, the W Single LLM baseline, which replaced all LLMs with the highest-performing LLM during \textit{pseudo-learning} (Mixtral-8x22b), showed lower performance compared to the full system.
This highlights the importance of using diverse LLMs.

The GET system also outperformed the system variant without \textit{pseudo-learning} on the UQ set.
This indicates the effectiveness of the full GET system for the UQ dataset, where prior examples are unavailable.
Especially, for this set, Gemini 1.5 Flash demonstrated the highest accuracy and F1 scores among the three graders, while the model used in the \textit{debate integration} phase is the different model, Mixtral-8x22b.
It can be said that \textit{pseudo-learning} outcomes enabled Mixtral-8x22b to improve overall performance during the debate phase.
This highlights the role of \textit{pseudo-learning} in enhancing the system adaptability across diverse datasets by helping the debating LLM identify the best grading candidates.

However, on the UA dataset, the system variant without\textit{pseudo-learning} achieved equivalent accuracy and slightly higher F1 scores.
To investigate the cause of this performance difference, we delved deeper into the three-class classification results by examining precision, recall, and F1-scores for each grading label (as defined in Eqs.\,\ref{eq-singlef1} and \ref{eq-precisionrecall}).
Among the three grading labels, the most noticeable difference was observed in the F1-score for the Incorrect label.
Specifically, the proposed system achieved an F1-score of 0.64705, while the partial variant (W/O Pseudo) achieved 0.68750.

Further analysis revealed that this drop in F1-score for the Incorrect class stemmed primarily from a decrease in precision.
While recall remained identical between the two systems, the proposed system exhibited a precision of 0.84615 compared to the W/O Pseudo baseline's perfect precision of 1.0.

The decline in F1-score for the Incorrect class can primarily be attributed to this decrease in precision.
This difference in precision arose because the proposed system misclassified two instances with a true label of Partially correct as Incorrect.
In contrast, the W/O Pseudo baseline did not misclassify any instances with true labels other than Incorrect as Incorrect.

Since the total number of Incorrect instances was only 21 (out of 375 total instances in the UA set), these two misclassifications had a significant impact on precision and, consequently, the F1 score.
This sensitivity highlights the challenges of dealing with imbalanced datasets and the critical role of precise classification in such cases.

\subsubsection{Grading Reason Generation Analysis}

The results of the LLM-based evaluation are shown in Table \ref{table-ablation_llmcritic}.
In this evaluation, our proposed system gets the highest scores on both the UA and UQ sets, indicating it has the best performance on generating grading reasons.

While the variant without \textit{pseudo-learning} partially surpassed the full system in grading label prediction, it underperformed in grading reason quality.
This suggests that \textit{pseudo-learning} has a greater impact on integrating grading reasons than on label prediction.
\begin{table}[!tb]
    \centering
    \caption{LLM-Based Evaluation results of GET and its partial variants}
    \label{table-ablation_llmcritic}
    \renewcommand{\arraystretch}{1.3}
    \begin{tabular}{lccc}
        \hline \textbf{} & UA              & UQ              & Avg.            \\ \hline
        Mixtral-8x22b    & 0.6969          & 0.6110          & 0.6539          \\ 
        Gemini 1.5 Flash & 0.6960          & 0.6319          & 0.6639          \\ 
        Llama-3-8B-it    & 0.5902          & 0.3709          & 0.4806          \\ 
        W/O Pseudo       & 0.6907          & 0.6228          & 0.6567          \\ 
        Only One LLM     & 0.6871          & 0.6284          & 0.6578          \\ 
        \textbf{GET}     & \textbf{0.7022} & \textbf{0.6360} & \textbf{0.6691} \\ \hline 
    \end{tabular}
\end{table}

Finally, for reference, Table \ref{table-ablation_bleu_rouge_bertscore} shows the results of the statistical evaluation of the quality of the grading reasons.
The proposed system achieved the highest BERTScore, especially on the UQ dataset, indicating that the grading reasons of the proposed system are close to the grand truth.

\begin{table}[!htb]
    \centering
    \caption{BLEU, ROUGE, and BERTScore of GET and its partial variants}
    \label{table-ablation_bleu_rouge_bertscore}
    \renewcommand{\arraystretch}{1.3}
    \begin{tabular}{lcccccc}
        \hline
        \textbf{}        & \multicolumn{2}{c}{\textbf{BLEU}} & \multicolumn{2}{c}{\textbf{ROUGE}} & \multicolumn{2}{c}{\textbf{BERTScore}}                                                       \\ \hline
                         & UA                                & UQ                                 & UA                                     & UQ              & UA              & UQ              \\ \hline
        Mixtral-8x22b    & 0.0248                            & 0.0164                             & 0.0567                                 & 0.0454          & 0.6620          & 0.6738          \\
        Gemini 1.5 Flash & 0.0134                            & 0.0154                             & 0.0299                                 & 0.0318          & 0.6604          & 0.6809          \\
        Llama3-8b-it     & 0.0256                            & 0.0179                             & 0.0507                                 & 0.0446          & 0.6189          & 0.6866          \\
        W/O Pseudo       & \textbf{0.0310}                   & 0.0277                             & \textbf{0.0720}                        & 0.0750          & \textbf{0.6973} & 0.7125          \\
        W Single LLM     & 0.0276                            & 0.0278                             & 0.0701                                 & \textbf{0.0801} & 0.6963          & 0.7133          \\
        \textbf{GET}     & 0.0288                            & \textbf{0.0289}                    & 0.0682                                 & 0.0772          & 0.6953          & \textbf{0.7171} \\ \hline
    \end{tabular}
\end{table}

\subsection{Limitations}\label{sec-limitations}
While this study demonstrates promising results, there are several limitations.
First, the quality of the output of the GET system needs further improvement, as we have discussed above.
Second, the system's performance was evaluated using only three specific LLMs, leaving its compatibility and effectiveness with other models untested.
Third, this study has not investigated actual students' perceptions and evaluations of the grading results generated by GET.
Addressing these limitations could provide valuable insights into the usability and accuracy of our system.

\section{Case Study}
\begin{figure}[!b]
    \centering
    \includegraphics[width=\textwidth]{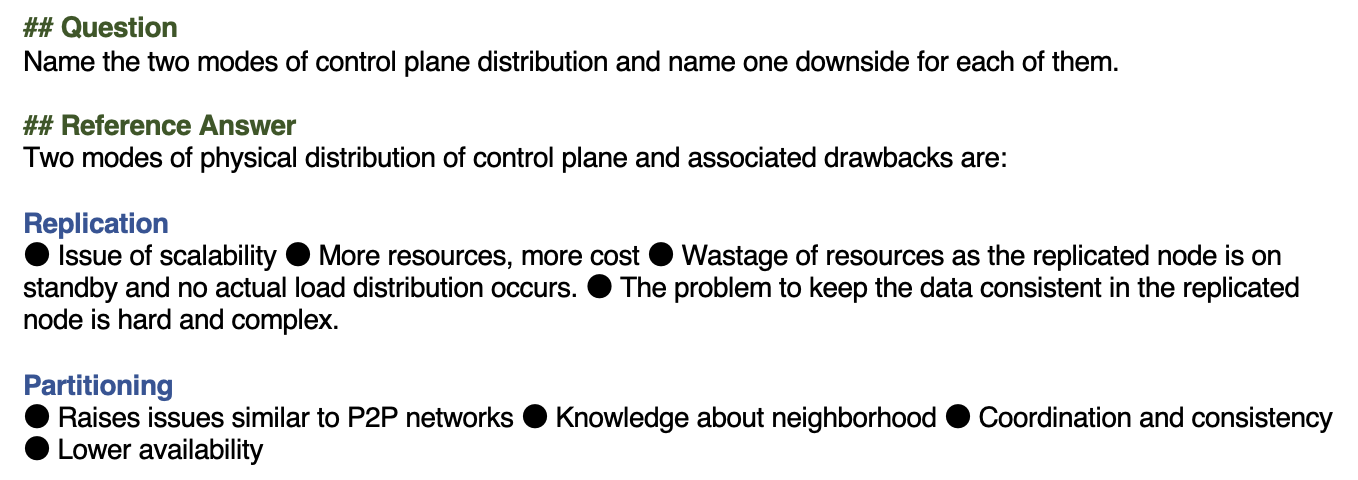}
    \caption{\textbf{Target Question and Reference Answer}}
    \label{fig-casestudytarget}
\end{figure}

This chapter supplements the statistical and LLM-based evaluations from the previous chapter with case studies of grading results.
We analyzed three student answers to a question from the UQ subset of the SAF dataset.
The target question and its reference answer are shown in Figure \ref{fig-casestudytarget}.
We present the GET system's outputs for three categories of student answers: Correct, Partially correct, and Incorrect.

\clearpage
\subsection{Correct Student Answer}

\begin{figure}[!b]
    \centering
    \includegraphics[width=\textwidth]{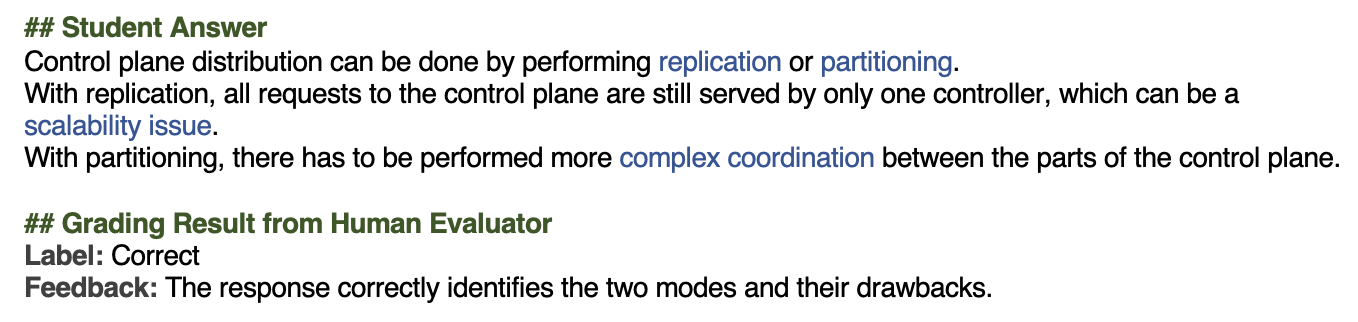}
    \caption{\textbf{Correct Student Answer}}
    \label{fig-sacorrect}
\end{figure}

\begin{figure}[!tb]
    \centering
    \includegraphics[width=\textwidth]{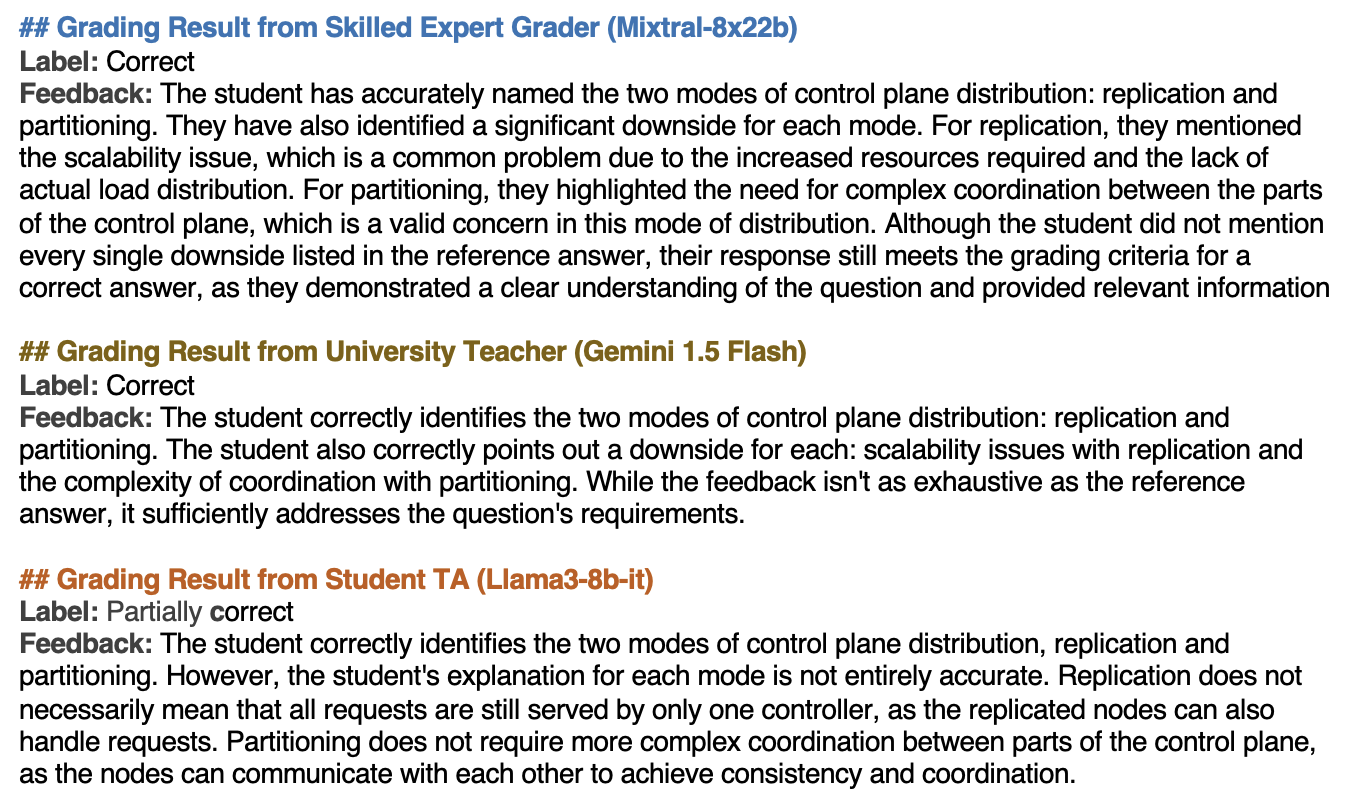}
    \caption{\textbf{Multi-LLM Grading Results (Correct Answer)}}
    \label{fig-mlgcorrect}
\end{figure}

This case examines a correct student answer (Figure \ref{fig-sacorrect}).
In the \textit{multi-LLM grading} phase (Figure \ref{fig-mlgcorrect}), the Skilled Expert Grader and University Teacher accurately assigned grades, while the Student TA mistakenly marked the answer as partially correct.

\begin{figure}[!tb]
    \centering
    \includegraphics[width=\textwidth]{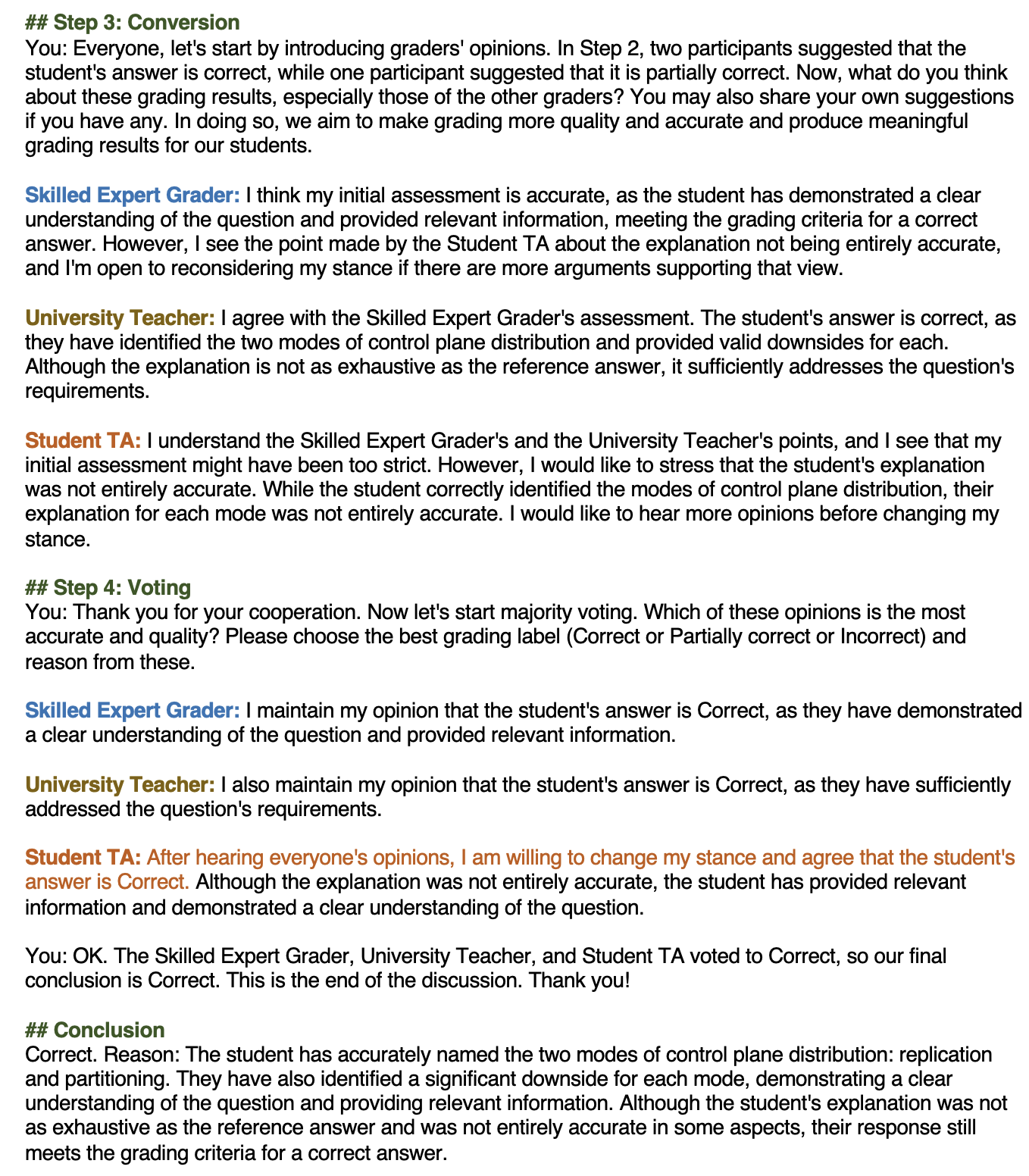}
    \caption{\textbf{Initial Debate Results (Correct Answer)}}
    \label{fig-idcorrect}
\end{figure}

Then, in the \textit{initial debate} phase (the first step of \textit{debate integration}), the Skilled Expert Grader facilitated a simulated debate (Figure \ref{fig-idcorrect}).
The debate began at Step 3 (Conversion), as Steps 1 (Ice Break) and 2 (Divergence) were pre-filled based on the outputs from \textit{multi-LLM grading} and \textit{pseudo-learning}.
During the debate, the Student TA revised the grade to align with the others.

\begin{figure}[!tb]
    \centering
    \includegraphics[width=\textwidth]{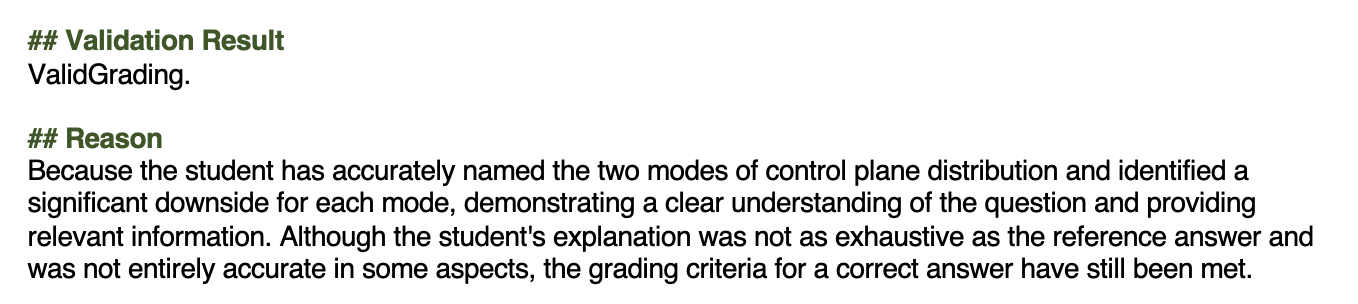}
    \caption{\textbf{Quality Validation Results (Correct Answer)}}
    \label{fig-qvcorrect}
\end{figure}

Finally, the \textit{quality validation} phase re-checked the integrated grading (Figure \ref{fig-qvcorrect}), confirming the \textit{initial debate} result and finalizing the GET system's output.

\clearpage
\subsection{Partially Correct Student Answer}

\begin{figure}[!b]
    \centering
    \includegraphics[width=\textwidth]{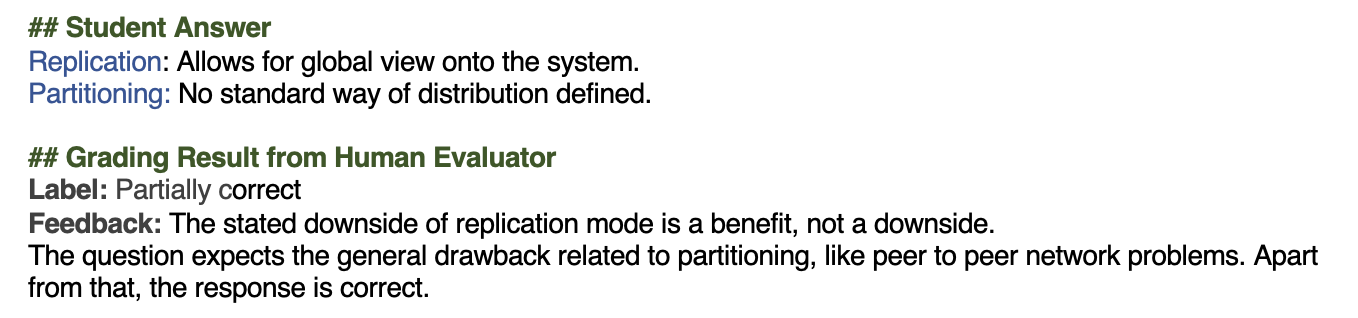}
    \caption{\textbf{Partially Correct Student Answer}}
    \label{fig-sapc}
\end{figure}

\begin{figure}[!htb]
    \centering
    \includegraphics[width=\textwidth]{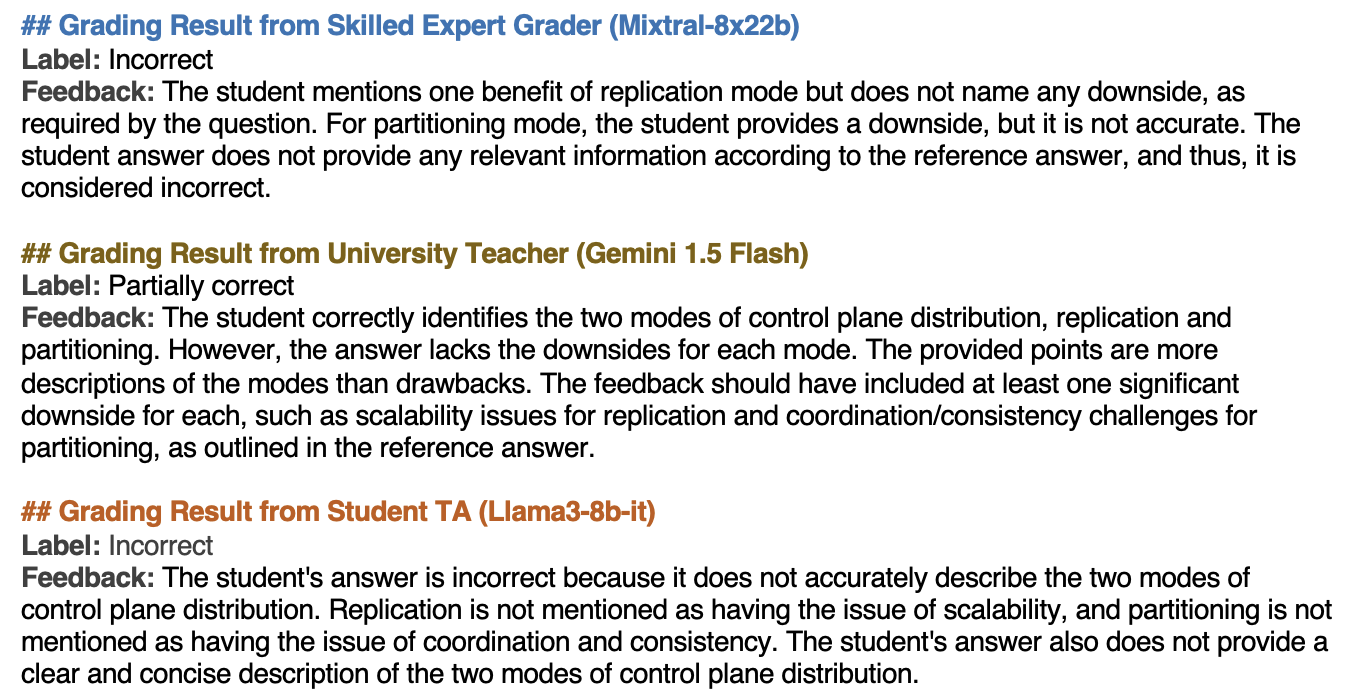}
    \caption{\textbf{Multi-LLM Grading Results (Partially Correct Answer)}}
    \label{fig-mlgpc}
\end{figure}

This case examines a partially correct student answer (Figure \ref{fig-sapc}).
In the \textit{multi-LLM grading} phase (Figure \ref{fig-mlgpc}), only the University Teacher's grade aligned with the human evaluator, marking the answer as partially correct.
The other two graders marked it as incorrect.
However, all graders identified the lack of drawbacks in the student's answer as a key issue.

\begin{figure}[!htb]
    \centering
    \centering
    \includegraphics[width=\textwidth]{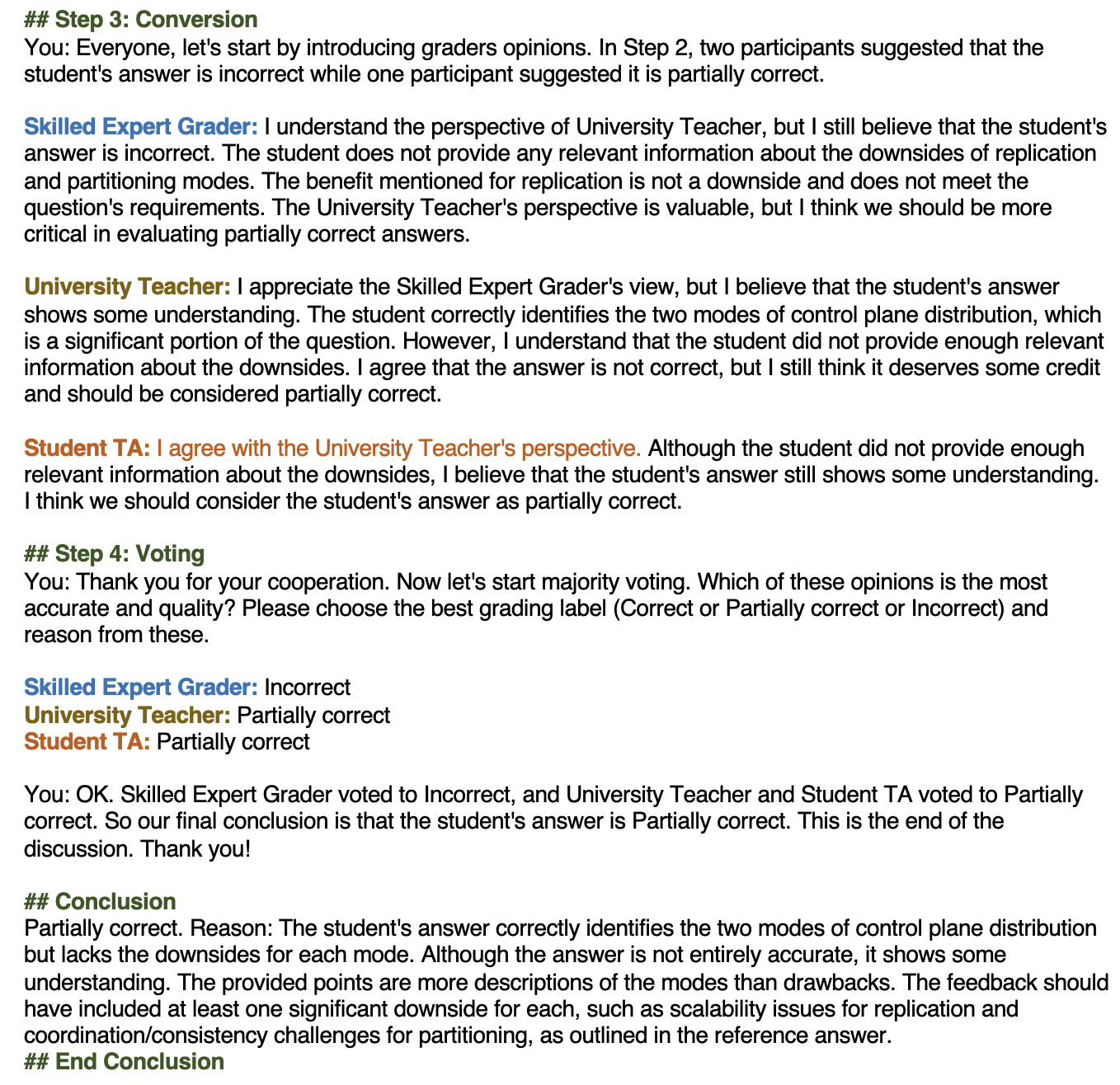}
    \caption{\textbf{Initial Debate Results (Partially Correct Answer)}}
    \label{fig-idpc}
\end{figure}

The \textit{initial debate} phase (Figure \ref{fig-idpc}) successfully resolved the discrepancy, with the Student TA changing the grade from incorrect to partially correct.
Despite the initial majority opinion, the GET system selected the most reasonable grading through the debate.

\begin{figure}[!htb]
    \centering
    \includegraphics[width=\textwidth]{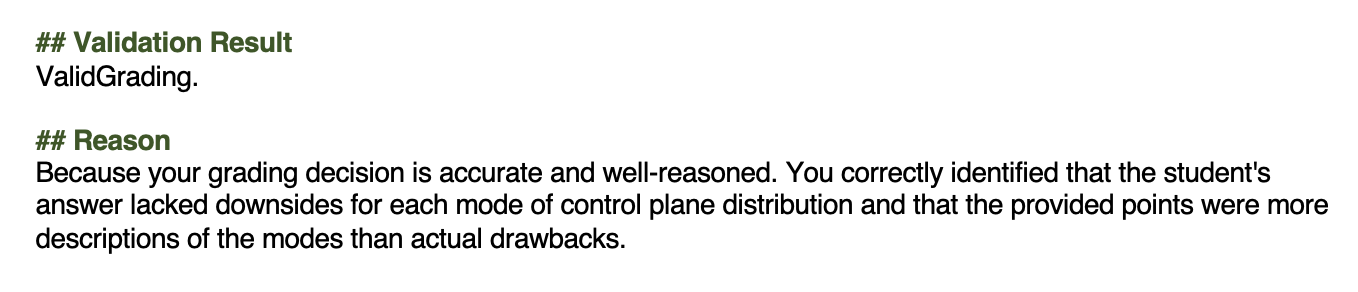}
    \caption{\textbf{Quality Validation Results (Partially Correct Answer)}}
    \label{fig-qvpc}
\end{figure}

The following \textit{quality validation} phase (Figure \ref{fig-qvpc}) confirmed the \textit{initial debate} results, finalizing the grade.

\clearpage

\subsection{Incorrect Student Answer}

\begin{figure}[!b]
    \centering
    \includegraphics[width=\textwidth]{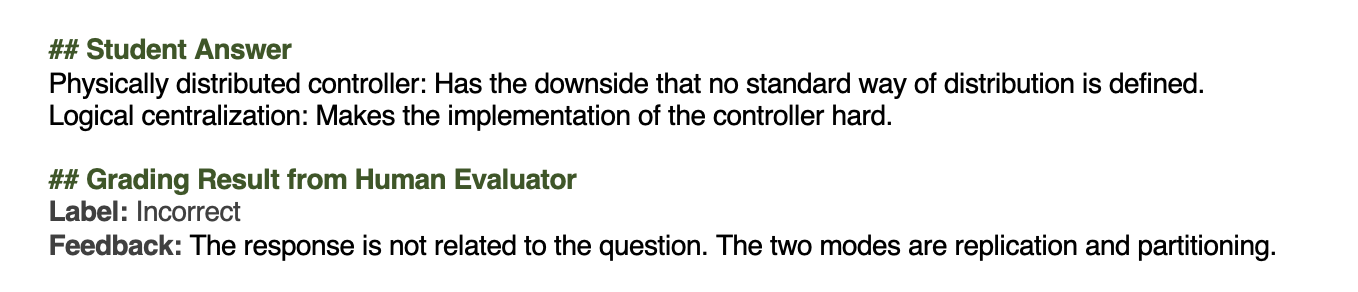}
    \caption{\textbf{Incorrect Student Answer}}
    \label{fig-sai}
\end{figure}

\begin{figure}[!htb]
    \centering
    \includegraphics[width=\textwidth]{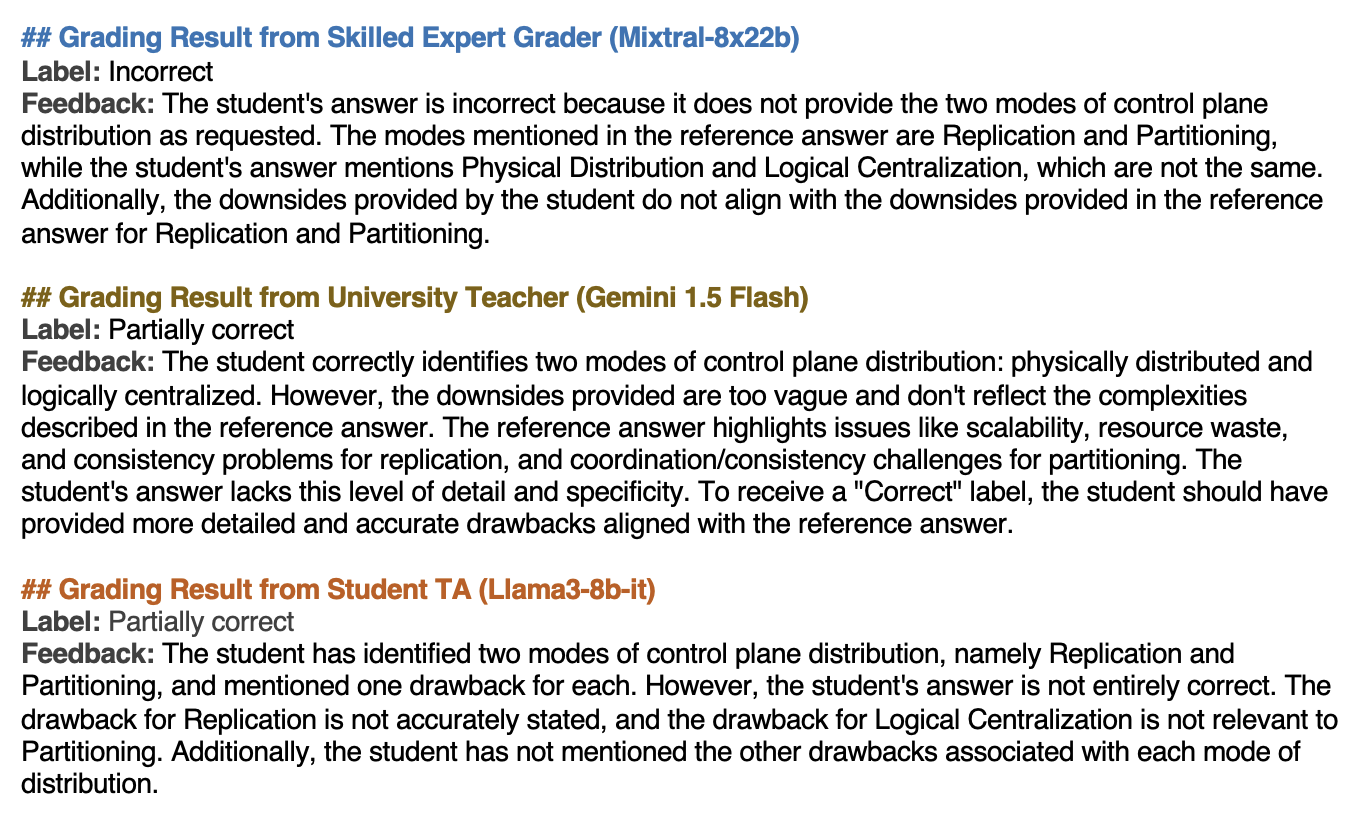}
    \caption{\textbf{Multi-LLM Grading Results (Incorrect Answer)}}
    \label{fig-mlgi}
\end{figure}

This case analyzes an incorrect student answer (Figure \ref{fig-sai}).
In the \textit{multi-LLM grading} phase (Figure \ref{fig-mlgi}), only the Skilled Expert Grader correctly identified the answer as incorrect.
The other two graders incorrectly marked it as partially correct with inconsistent reasoning.

\begin{figure}[!htb]
    \centering
    \includegraphics[width=\textwidth]{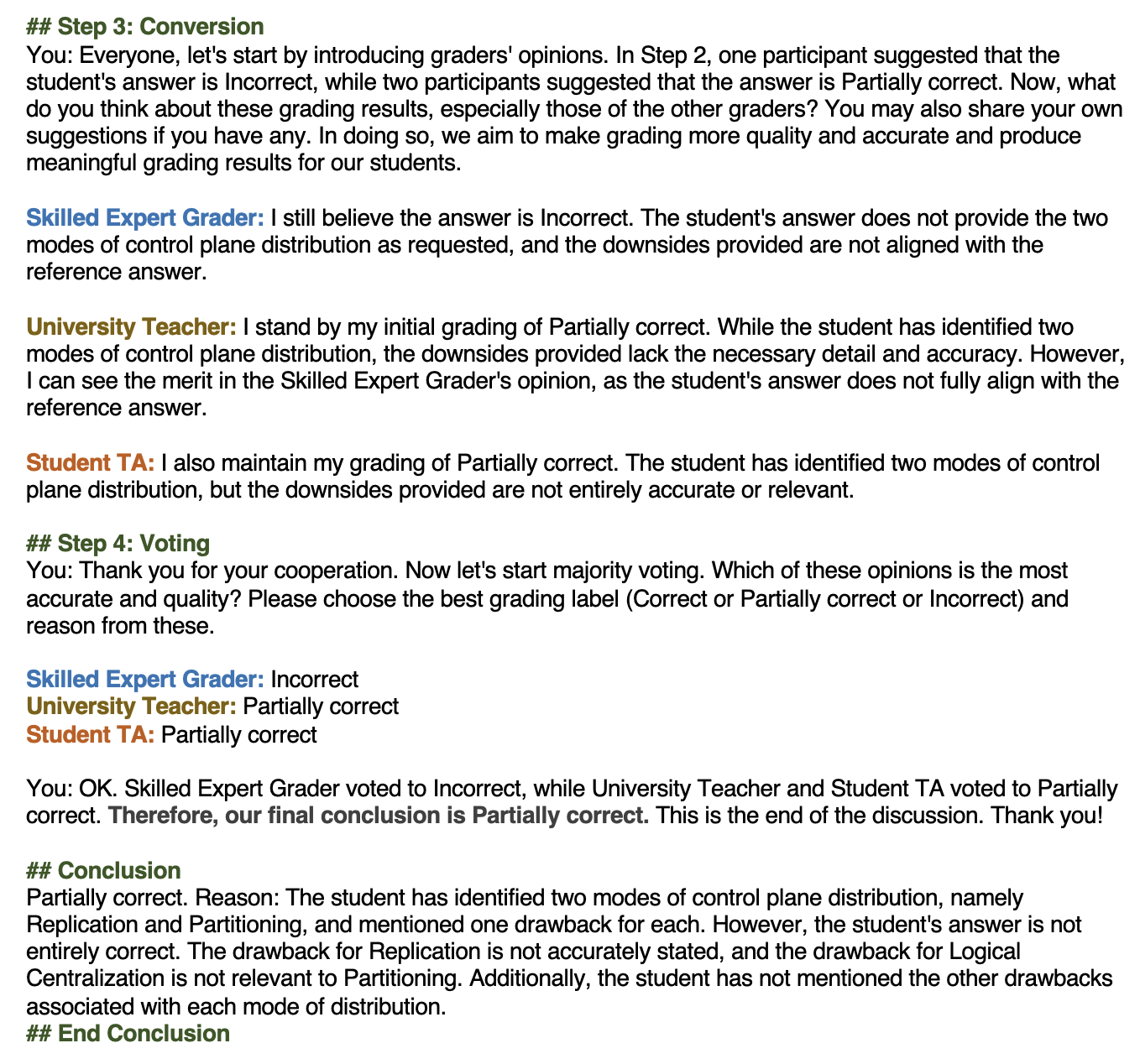}
    \caption{\textbf{Initial Debate Results (Incorrect Answer)}}
    \label{fig-idi}
\end{figure}

The \textit{initial debate} phase (Figure \ref{fig-idi}) failed to reach a valid conclusion due to the inaccurate initial grades from the University Teacher and Student TA.

\begin{figure}[!htb]
    \centering
    \includegraphics[width=\textwidth]{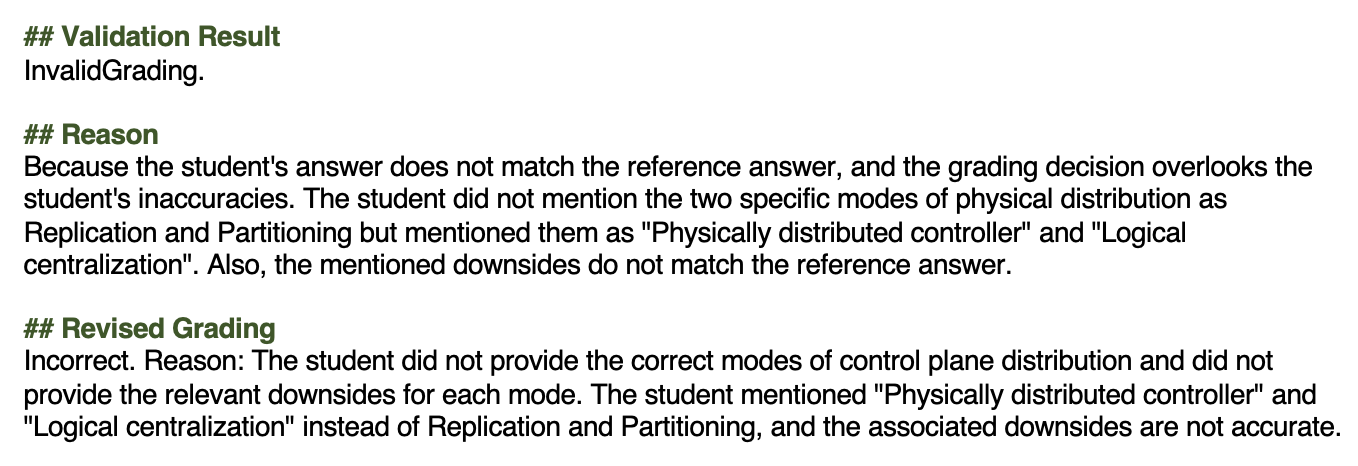}
    \caption{\textbf{Quality Validation Results (Incorrect Answer)}}
    \label{fig-qvi}
\end{figure}

\begin{figure}[!htb]
    \centering
    \includegraphics[width=\textwidth]{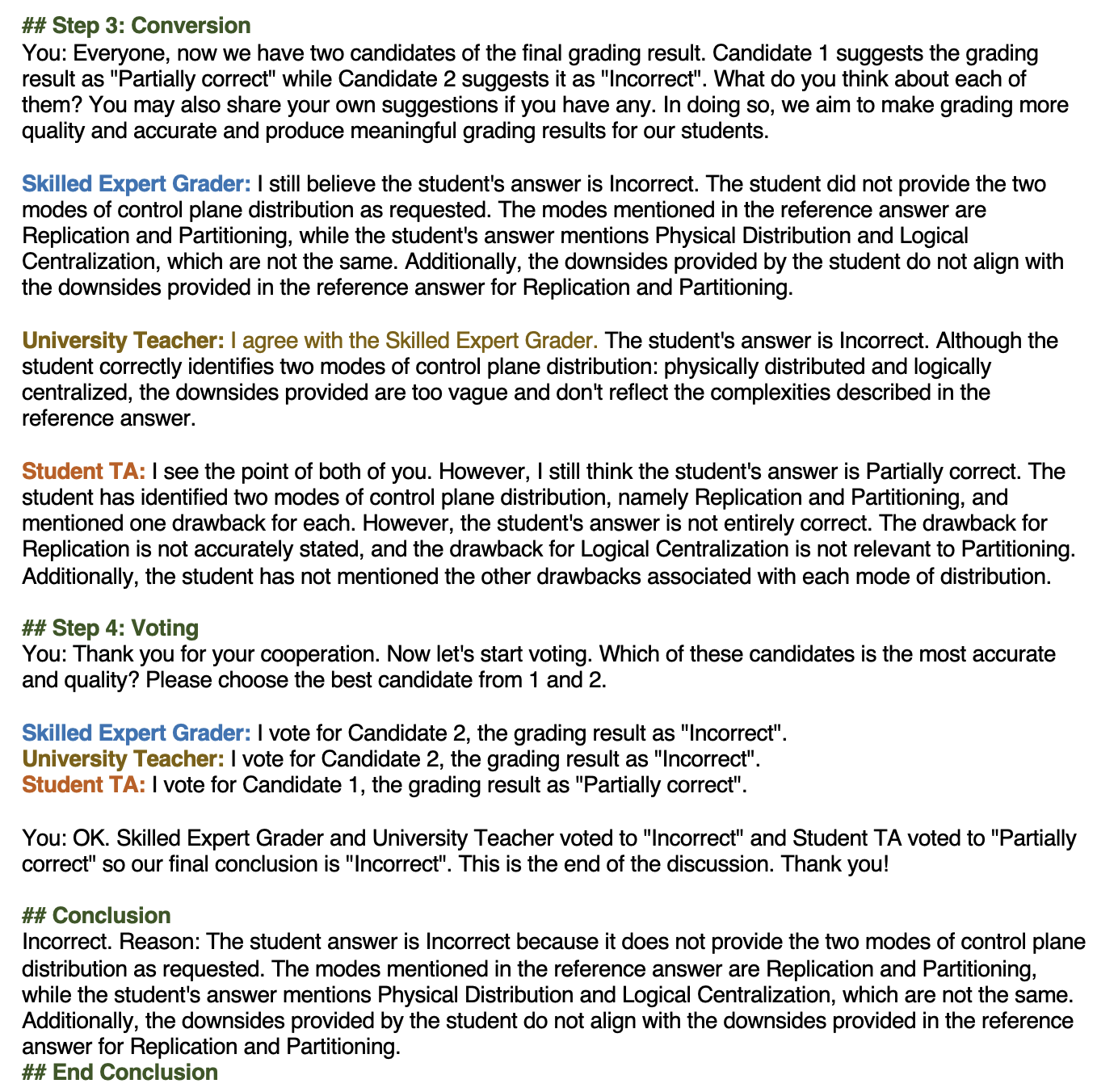}
    \caption{\textbf{Debate Retry Results (Incorrect Answer)}}
    \label{fig-dri}
\end{figure}

However, during \textit{quality validation} (Figure \ref{fig-qvi}), the system successfully identified the invalid debate result and suggested a revised grading.
This triggered the \textit{debate retry} phase (Figure \ref{fig-dri}).
In this phase, graders reconsider their conclusions via additional debate.
Specifically, the system presents the LLMs with two candidates: Candidate 1, the conclusion from the \textit{initial debate}, and Candidate 2, the revised grading suggested by the \textit{quality validation}.
The graders then select one of these candidates.
In this case, the University Teacher revised the opinion, leading to the appropriate grading.

\clearpage
\section{Conclusion}\label{sec-conclusion}
In this study, we proposed a novel framework to collaborate multiple LLMs: \textbf{Ensemble ToT} and an automatic grading system, \textbf{Graders by Ensemble ToT (GET)}.
In the Ensemble ToT framework, LLMs performance tendencies on a specific task are identified, which is inspired by the ensemble learning method.
Then multiple LLMs separately generate the candidate solutions for the target task as in the Tree-of-Thought framework.
Finally, their outputs are combined based on the performance tendencies identified before.

The GET system is an automatic grading system built on the Ensemble ToT framework.
It achieves high-accuracy grading by integrating multiple LLMs via the following three steps: \textit{pseudo-learning}, \textit{multi-LLM grading} and \textit{debate integration}.
During the \textit{pseudo-learning} phase, the characteristics of each LLM are analyzed.
In the subsequent \textit{multi-LLM grading} phase, multiple LLMs individually grade the student answers.
Then, in \textit{debate integration} phase, the final grading results are determined through a discussion process.
This approach complements the weaknesses of individual LLMs, enabling accurate and balanced grading and its detailed explanation.

The experimental results demonstrate that the GET system significantly outperformed several baseline methods in grading accuracy.
Specifically, our system achieved the maximum grading label prediction accuracy of 77.87\%, representing a 6.94\% improvement over the state-of-the-art method. Additionally, the GET system attained the maximum macro F1 score of 0.7128, reflecting an 8.13\% increase.
In the LLM-based evaluation, which measures the generation of valid and coherent grading reasons, the GET system also surpassed ASAS-F-RAG, achieving a 4.41\% average rating score improvement.
These findings highlight the GET system's capability to produce both accurate and explainable grading outcomes.

\section*{Acknowledgment}
Part of this research was supported by Grant-in-Aid for Scientific Research (23K28094).

\bibliographystyle{unsrtnat}
\bibliography{references}

\begin{thebibliography}{32}
\providecommand{\natexlab}[1]{#1}
\providecommand{\url}[1]{\texttt{#1}}
\expandafter\ifx\csname urlstyle\endcsname\relax
  \providecommand{\doi}[1]{doi: #1}\else
  \providecommand{\doi}{doi: \begingroup \urlstyle{rm}\Url}\fi

\bibitem[Chan and Hu(2023)]{voicesonai}
Cecilia Ka~Yuk Chan and Wenjie Hu.
\newblock Students'voices on generative ai: perceptions, benefits, and
  challenges in higher education.
\newblock \emph{Int. J. Educ. Technol. High. Educ.}, 20\penalty0 (1):\penalty0
  43, 2023.
\newblock \doi{10.1186/s41239-023-00411-8}.
\newblock URL \url{https://doi.org/10.1186/s41239-023-00411-8}.

\bibitem[del Gobbo et~al.(2023)del Gobbo, Guarino, Cafarelli, and
  Grilli]{gradeaid}
Emiliano del Gobbo, Alfonso Guarino, Barbara Cafarelli, and Luca Grilli.
\newblock Gradeaid: a framework for automatic short answers grading in
  educational contexts - design, implementation and evaluation.
\newblock \emph{Knowl. Inf. Syst.}, 65\penalty0 (10):\penalty0 4295--4334,
  2023.
\newblock \doi{10.1007/S10115-023-01892-9}.
\newblock URL \url{https://doi.org/10.1007/s10115-023-01892-9}.

\bibitem[Huang et~al.(2023)Huang, Lee, and Kwon]{direct}
Jin{-}Xia Huang, Yohan Lee, and Oh{-}Woog Kwon.
\newblock {DIRECT:} toward dialogue-based reading comprehension tutoring.
\newblock \emph{{IEEE} Access}, 11:\penalty0 8978--8987, 2023.
\newblock \doi{10.1109/ACCESS.2022.3233224}.
\newblock URL \url{https://doi.org/10.1109/ACCESS.2022.3233224}.

\bibitem[Ormerod(2022)]{ensembleoflm}
Christopher Ormerod.
\newblock Short-answer scoring with ensembles of pretrained language models.
\newblock \emph{arXiv preprint arXiv:2202.11558}, 2022.

\bibitem[Zhang et~al.(2024)Zhang, Sun, Chen, Pfister, Zhang, and Arik]{CoA}
Yusen Zhang, Ruoxi Sun, Yanfei Chen, Tomas Pfister, Rui Zhang, and
  Sercan~{\"{O}}. Arik.
\newblock Chain of agents: Large language models collaborating on long-context
  tasks.
\newblock \emph{CoRR}, abs/2406.02818, 2024.
\newblock \doi{10.48550/ARXIV.2406.02818}.
\newblock URL \url{https://doi.org/10.48550/arXiv.2406.02818}.

\bibitem[Yang et~al.(2023)Yang, Li, Zhou, Xiao, Fang, and Zhang]{yang2023one}
Han Yang, Mingchen Li, Huixue Zhou, Yongkang Xiao, Qian Fang, and Rui Zhang.
\newblock One llm is not enough: Harnessing the power of ensemble learning for
  medical question answering.
\newblock \emph{medRxiv}, 2023.

\bibitem[Xiong et~al.(2023)Xiong, Ding, Cao, Liu, and Qin]{ford}
Kai Xiong, Xiao Ding, Yixin Cao, Ting Liu, and Bing Qin.
\newblock Examining inter-consistency of large language models collaboration:
  An in-depth analysis via debate.
\newblock In \emph{Findings of the Association for Computational Linguistics:
  {EMNLP} 2023, Singapore, December 6-10, 2023}, pages 7572--7590. Association
  for Computational Linguistics, 2023.
\newblock \doi{10.18653/V1/2023.FINDINGS-EMNLP.508}.
\newblock URL \url{https://doi.org/10.18653/v1/2023.findings-emnlp.508}.

\bibitem[Chang and Ginter(2024)]{finnishgrading}
Li-Hsin Chang and Filip Ginter.
\newblock Automatic short answer grading for finnish with chatgpt.
\newblock In \emph{Proc. AAAI Conf. Artif. Intell.}, volume~38, pages
  23173--23181, 2024.

\bibitem[Gobrecht et~al.(2024)Gobrecht, Tuma, M{\"{o}}ller, Z{\"{o}}ller,
  Zakhvatkin, et~al.]{beyondhumansubjectivity}
Alexandra Gobrecht, Felix Tuma, Moritz M{\"{o}}ller, Thomas Z{\"{o}}ller, Mark
  Zakhvatkin, et~al.
\newblock Beyond human subjectivity and error: a novel {AI} grading system.
\newblock \emph{arXiv preprint arXiv:2405.04323}, 2024.

\bibitem[Ito and Ma(2025)]{icetc}
Yuki Ito and Qiang Ma.
\newblock Supporting student self-learning using generative ai.
\newblock In \emph{Proc. 16th Int. Conf. Educ. Technol. Comput. 2024}, ICETC
  '24, page 97–103, New York, NY, USA, 2025. Association for Computing
  Machinery.
\newblock ISBN 9798400717819.
\newblock \doi{10.1145/3702163.3702177}.
\newblock URL \url{https://doi.org/10.1145/3702163.3702177}.

\bibitem[Lee et~al.(2024)Lee, Latif, Wu, Liu, and Zhai]{cotscoring}
Gyeong{-}Geon Lee, Ehsan Latif, Xuansheng Wu, Ninghao Liu, and Xiaoming Zhai.
\newblock Applying large language models and chain-of-thought for automatic
  scoring.
\newblock \emph{Comput. Educ. Artif. Intell.}, 6:\penalty0 100213, 2024.
\newblock \doi{10.1016/J.CAEAI.2024.100213}.
\newblock URL \url{https://doi.org/10.1016/j.caeai.2024.100213}.

\bibitem[Jiang and Bosch(2024)]{sasgpt4}
Lan Jiang and Nigel Bosch.
\newblock Short answer scoring with {GPT-4}.
\newblock In \emph{Proc. 11th {ACM} Conf. Learning @ Scale (L@S 2024)}, pages
  438--442. {ACM}, 2024.
\newblock \doi{10.1145/3657604.3664685}.
\newblock URL \url{https://doi.org/10.1145/3657604.3664685}.

\bibitem[Matelsky et~al.(2023)Matelsky, Parodi, Liu, Lange, and
  Kording]{freetext}
Jordan~K. Matelsky, Felipe Parodi, Tony Liu, Richard~D. Lange, and Konrad~P.
  Kording.
\newblock A large language model-assisted education tool to provide feedback on
  open-ended responses.
\newblock \emph{arXiv preprint arXiv:2308.02439}, 2023.

\bibitem[Fateen et~al.(2024)Fateen, Wang, and Mine]{beyondscores}
Menna Fateen, Bo~Wang, and Tsunenori Mine.
\newblock Beyond scores: A modular rag-based system for automatic short answer
  scoring with feedback.
\newblock \emph{IEEE Access}, 2024.

\bibitem[Filighera et~al.(2022)Filighera, Parihar, Steuer, Meuser, and
  Ochs]{safdataset}
Anna Filighera, Siddharth Parihar, Tim Steuer, Tobias Meuser, and Sebastian
  Ochs.
\newblock Your answer is incorrect... would you like to know why? introducing a
  bilingual short answer feedback dataset.
\newblock In \emph{Proc. 60th Annu. Meeting of the Assoc. Comput. Linguistics
  (Vol. 1: Long Papers)}, pages 8577--8591, Dublin, Ireland, May 2022.
  Association for Computational Linguistics.
\newblock \doi{10.18653/v1/2022.acl-long.587}.
\newblock URL \url{https://aclanthology.org/2022.acl-long.587}.

\bibitem[Aggarwal et~al.(2024)Aggarwal, Bhattacharyya, and
  Raman]{iunderstandwhy}
Dishank Aggarwal, Pushpak Bhattacharyya, and Bhaskaran Raman.
\newblock "i understand why {I} got this grade": Automatic short answer grading
  with feedback.
\newblock \emph{arXiv preprint arXiv:2407.12818}, 2024.

\bibitem[Yao et~al.(2023)Yao, Yu, Zhao, Shafran, Griffiths, et~al.]{ToT}
Shunyu Yao, Dian Yu, Jeffrey Zhao, Izhak Shafran, Tom Griffiths, et~al.
\newblock Tree of thoughts: Deliberate problem solving with large language
  models.
\newblock In \emph{Advances in Neural Inf. Process. Syst. 36: Annu. Conf.
  Neural Inf. Process. Syst. (NeurIPS 2023)}, 2023.
\newblock URL
  \url{http://papers.nips.cc/paper\_files/paper/2023/hash/271db9922b8d1f4dd7aaef84ed5ac703-Abstract-Conference.html}.

\bibitem[Ibomoiye and Sun(2022)]{ensemblelearning}
Domor~Mienye Ibomoiye and Yanxia Sun.
\newblock A survey of ensemble learning: Concepts, algorithms, applications,
  and prospects.
\newblock \emph{{IEEE} Access}, 10:\penalty0 99129--99149, 2022.
\newblock \doi{10.1109/ACCESS.2022.3207287}.
\newblock URL \url{https://doi.org/10.1109/ACCESS.2022.3207287}.

\bibitem[Wolpert(1992)]{stacking}
David~H. Wolpert.
\newblock Stacked generalization.
\newblock \emph{Neural Networks}, 5\penalty0 (2):\penalty0 241--259, 1992.
\newblock \doi{10.1016/S0893-6080(05)80023-1}.
\newblock URL \url{https://doi.org/10.1016/S0893-6080(05)80023-1}.

\bibitem[Pedregosa et~al.(2011)Pedregosa, Varoquaux, Gramfort, Michel, Thirion,
  Grisel, Blondel, Prettenhofer, Weiss, Dubourg, Vanderplas, Passos,
  Cournapeau, Brucher, Perrot, and Duchesnay]{scikit-learn}
F.~Pedregosa, G.~Varoquaux, A.~Gramfort, V.~Michel, B.~Thirion, O.~Grisel,
  M.~Blondel, P.~Prettenhofer, R.~Weiss, V.~Dubourg, J.~Vanderplas, A.~Passos,
  D.~Cournapeau, M.~Brucher, M.~Perrot, and E.~Duchesnay.
\newblock Scikit-learn: Machine learning in {P}ython.
\newblock \emph{J. Mach. Learn. Res.}, 12:\penalty0 2825--2830, 2011.

\bibitem[Khattab and Zaharia(2020)]{colbert}
Omar Khattab and Matei Zaharia.
\newblock Colbert: Efficient and effective passage search via contextualized
  late interaction over {BERT}.
\newblock In \emph{Proc. 43rd Int. {ACM} {SIGIR} Conf. Res. Develop. Inf.
  Retrieval, {SIGIR} 2020}, pages 39--48. {ACM}, 2020.
\newblock \doi{10.1145/3397271.3401075}.
\newblock URL \url{https://doi.org/10.1145/3397271.3401075}.

\bibitem[Dong et~al.(2024)Dong, Ding, and Ito]{llmfacilitator}
Yihan Dong, Shiyao Ding, and Takayuki Ito.
\newblock An automated multi-phase facilitation agent based on {LLM}.
\newblock \emph{{IEICE} Trans. Inf. Syst.}, 107\penalty0 (4):\penalty0
  426--433, 2024.
\newblock \doi{10.1587/TRANSINF.2023IHP0011}.
\newblock URL \url{https://doi.org/10.1587/transinf.2023ihp0011}.

\bibitem[Papineni et~al.(2002)Papineni, Roukos, Ward, and Zhu]{bleu}
Kishore Papineni, Salim Roukos, Todd Ward, and Wei{-}Jing Zhu.
\newblock Bleu: a method for automatic evaluation of machine translation.
\newblock In \emph{Proc. 40th Annu. Meeting of the Assoc. Comput. Linguistics},
  pages 311--318. {ACL}, 2002.
\newblock \doi{10.3115/1073083.1073135}.
\newblock URL \url{https://aclanthology.org/P02-1040/}.

\bibitem[Lin(2004)]{rouge}
Chin-Yew Lin.
\newblock {ROUGE}: A package for automatic evaluation of summaries.
\newblock In \emph{Text Summarization Branches Out}, pages 74--81, Barcelona,
  Spain, July 2004. Assoc. Comput. Linguistics.
\newblock URL \url{https://aclanthology.org/W04-1013}.

\bibitem[Zhang et~al.(2020)Zhang, Kishore, Wu, Weinberger, and
  Artzi]{bertscore}
Tianyi Zhang, Varsha Kishore, Felix Wu, Kilian~Q. Weinberger, and Yoav Artzi.
\newblock Bertscore: Evaluating text generation with {BERT}.
\newblock In \emph{8th Int. Conf. Learn. Representations (ICLR 2020)}.
  OpenReview.net, 2020.
\newblock URL \url{https://openreview.net/forum?id=SkeHuCVFDr}.

\bibitem[Jurenka et~al.(2024)Jurenka, Kunesch, McKee, Gillick, Zhu,
  et~al.]{towardsresponsible}
Irina Jurenka, Markus Kunesch, Kevin~R. McKee, Daniel Gillick, Shaojian Zhu,
  et~al.
\newblock Towards responsible development of generative {AI} for education: An
  evaluation-driven approach.
\newblock \emph{arXiv preprint arXiv:2407.12687}, 2024.

\bibitem[Reid et~al.(2024)Reid, Savinov, Teplyashin, Lepikhin, Lillicrap,
  et~al.]{gemini}
Machel Reid, Nikolay Savinov, Denis Teplyashin, Dmitry Lepikhin, Timothy~P.
  Lillicrap, et~al.
\newblock Gemini 1.5: Unlocking multimodal understanding across millions of
  tokens of context.
\newblock \emph{arXiv preprint arXiv:2403.05530}, 2024.

\bibitem[{Meta}()]{llamabench}
{Meta}.
\newblock Meta-llama-3-8b.
\newblock \url{https://huggingface.co/} \allowbreak
  {meta-llama/Meta-Llama-3-8B}.
\newblock Accessed: Dec 26, 2024.

\bibitem[{Google DeepMind}()]{geminibenck}
{Google DeepMind}.
\newblock Gemini.
\newblock \url{https://deepmind.google/technologies/gemini/}.
\newblock Accessed: Dec 26, 2024.

\bibitem[{Mistral AI}()]{mixtralbench}
{Mistral AI}.
\newblock Mixtral 8x22b.
\newblock \url{https://mistral.ai/news/mixtral-8x22b/}.
\newblock Accessed: Dec 26, 2024.

\bibitem[AI@Meta(2024)]{llama3modelcard}
AI@Meta.
\newblock Llama 3 model card.
\newblock \url{https://github.com/meta-llama/llama3/blob/main/MODEL_CARD.md},
  2024.

\bibitem[Jiang et~al.(2024)Jiang, Sablayrolles, Roux, Mensch, Savary,
  et~al.]{mixtral}
Albert~Q. Jiang, Alexandre Sablayrolles, Antoine Roux, Arthur Mensch, Blanche
  Savary, et~al.
\newblock Mixtral of experts.
\newblock \emph{arXiv preprint arXiv:2401.04088}, 2024.

\end{thebibliography}

\end{document}